\documentclass[12pt]{iopart}

\usepackage{harvard}
\citationmode{abbr}
\usepackage{hyperref}
\usepackage{booktabs}
\usepackage{graphicx}

\expandafter\let\csname equation*\endcsname\relax
\expandafter\let\csname endequation*\endcsname\relax
\usepackage{ulem}
\usepackage{amsmath}
\usepackage{gensymb}
\usepackage{fancyhdr} 
\usepackage{lineno} 
\usepackage{csvsimple} 
\usepackage{lscape} 
\usepackage{multicol}
\usepackage{multirow}
\usepackage[table]{xcolor} 

\begin{document}

\title{Annual field-scale maps of tall and short crops at the global scale using GEDI and Sentinel-2}

 \author{Stefania Di Tommaso\textsuperscript{1}, Sherrie Wang\textsuperscript{2,3}, Vivek Vajipey\textsuperscript{1}, Noel Gorelick\textsuperscript{4,5}, Rob Strey\textsuperscript{6}, David B. Lobell\textsuperscript{1}{*}}

\address{\textsuperscript{1} Department of Earth System Science and Center on Food Security and the Environment, Stanford University}
\address{\textsuperscript{2} Goldman School of Public Policy, University of California, Berkeley}
\address{\textsuperscript{3} MIT}
\address{\textsuperscript{4} Google, Zürich, Switzerland}
\address{\textsuperscript{5} Remote Sensing Laboratories, University of Zürich, Zürich, Switzerland}
\address{\textsuperscript{6} Progressive Environmental \& Agricultural Technologies, 10435 Berlin, Germany}

\ead{dlobell@stanford.edu}


\begin{abstract}

Crop type maps are critical for tracking agricultural land use and estimating crop production. Remote sensing has proven an efficient and reliable tool for creating these maps in regions with abundant ground labels for model training, yet these labels remain difficult to obtain in many regions and years. NASA's Global Ecosystem Dynamics Investigation (GEDI) spaceborne lidar instrument, originally designed for forest monitoring, has shown promise for distinguishing tall and short crops. In the current study, we leverage GEDI to develop wall-to-wall maps of short vs tall crops on a global scale at 10 m resolution for 2019-2021. Specifically, we show that (i) GEDI returns can reliably be classified into tall and short crops after removing shots with extreme view angles or topographic slope, (ii) the frequency of tall crops over time can be used to identify months when tall crops are at their peak height, and (iii) GEDI shots in these months can then be used to train random forest models that use Sentinel-2 time series to accurately predict short vs. tall crops. Independent reference data from around the world are then used to evaluate these GEDI-S2 maps. We find that GEDI-S2 performed nearly as well as models trained on thousands of local reference training points, with accuracies of at least 87\% and often above 90\% throughout the Americas, Europe, and East Asia. Systematic underestimation of tall crop area was observed in regions where crops frequently exhibit low biomass, namely Africa and South Asia, and further work is needed in these systems. Although the GEDI-S2 approach only differentiates tall from short crops, in many landscapes this distinction goes a long way toward mapping the main individual crop types (e.g., maize vs. soy, sugarcane vs. rice). The combination of GEDI and Sentinel-2 thus presents a very promising path towards global crop mapping with minimal reliance on ground data.

\end{abstract}

\section{Introduction}\label{intro}

Farmer livelihoods and food production are affected by myriad ongoing changes in climate, markets, and policies. Accurate data on cropping systems are essential to monitor and understand the effects of these changes, yet such data are often lacking. One major aspect of cropping systems are the crops that farmers choose to plant, which typically change from season to season as farmers rotate crops or shift into new crops \cite{begue2018remote}. Information on crop choice is helpful for various applications, including modeling land use decisions, mapping yield variations, and forecasting regional food production.

Given the widespread demand for crop type information, maps of crop types have been developed from a variety of sources and at a range of spatial and temporal resolutions \cite{kim2021review,nakalembe2021review}. In a small number of countries, such as the United States \cite{boryan2011monitoring}, Canada \cite{aafc}, and France \cite{france}, detailed crop type maps at the field scale are publicly available for each growing season, based either on farmer surveys or a combination of ground and satellite sources. In most countries, however, timely data is much harder to obtain. Several gridded datasets with global coverage have been developed, but these are often based on census data more than a decade old \cite{kim2021review}. Given the dynamic nature of agriculture, including evidence of rapid cropland expansion in some regions and cropland abandonment in others \cite{potapov2022global}, decades old data are insufficient for many uses. Moreover, global products typically have a resolution of 10 km or coarser \cite{kim2021review}, which limits their utility for applications requiring field-scale data.

As a result, there is a continued need for improved approaches to mapping crop types \cite{kim2021review,nakalembe2021review}. This need is recognized, for example, by the new WorldCereal effort that aims to create global, annual maps for wheat and maize at 10 m resolution (https://esa-worldcereal.org/). Remote sensing offers clear advantages for large-scale crop type mapping, with proven success in many local or national scale studies \cite{lobell2003remote,boryan2011monitoring,you202110m,han2021asiaricemap10m,song2021massive,blickensdorfer2022mapping}. Yet a major challenge remains that models require large amounts of training data, and models trained in one region for a single season often do not transfer well to other regions or seasons. One sensible way to address this challenge, as in the WorldCereal project, is to invest in large amounts of field data collection around the world, so that models can be locally trained anywhere. Other efforts have focused on developing models that are better able to maintain performance in years or locations outside of their training domain \cite{kluger2021two,lin2022early,luo2022developing}. 

\begin{figure}
	\centering
	\includegraphics[width=1.0\linewidth]{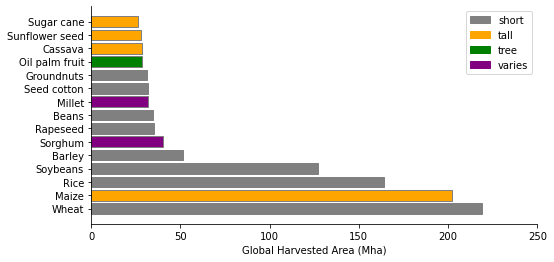}
    \caption{
    Most commonly grown crops in the world based on FAO crop statistics for 2019, color coded by crop heights base on U.S. National Plan Germplasm System. Crops are considered tall if most common varieties exceed 2 m in peak height.
    Data source: \url{http://www.fao.org/faostat} for global crop areas, \url{https://npgsweb.ars-grin.gov} for crop heights.
    }
    \label{fig:faostats}
\end{figure}

A third, complementary approach has been to seek training data derived without the need for field data collection. In recent work \cite{di2021combining}, we demonstrated the promise of one such source of data -- lidar measurements acquired by the Global Ecosystem Dynamics Investigation (GEDI) \cite{dubayah2020global}. GEDI lidar returns provide information on canopy heights with a nominal spatial resolution of 25 m and a vertical precision of roughly 50 cm \cite{dubayah2020global}. Although many crops have similar heights, some of the key commodity crops grown throughout the world, especially maize and sugarcane, are typically 1 m taller than other common crops such as wheat, rice, or soybean (Figure \ref{fig:faostats})
. Indeed, in many landscapes the two main crops are one tall crop and one short crop (e.g. maize and soybean, or sugarcane and rice), such that the ability to discern tall from short crops goes a long way toward mapping individual crops. GEDI alone, however, only samples a very small fraction of the landscape, so therefore rather than use GEDI directly, \citeasnoun{di2021combining} use GEDI to train a random forest model that predicts crop height class based on Sentinel-2 (S2) optical data. This combined GEDI-S2 approach was found to map crop types nearly as well as a model trained on thousands of local ground training points in the US, France, and China.  

Here we extend the analysis of \citeasnoun{di2021combining} to evaluate the performance of a GEDI-S2 approach across a broader range of conditions in both the cropping system and the satellite sensor configuration. Specifically, we implement GEDI-S2 on a global scale for 2019-2021, expanding well beyond the original study of three regions in a single year. Over this time period, not only did the climatic conditions experienced by crops vary, but the orbiting altitude and view angle of the GEDI instrument also varied significantly. Given the relatively small signal ($\sim$1 m) being used to discriminate crop types, these sensor shifts could represent an important source of noise. Similarly, extending beyond the relatively high yielding systems of U.S., France, and China presents a challenge to the GEDI-S2 approach, given that lower yielding systems will have fewer photons reflected off of leaves near the top of the canopy, and smaller fields will be more likely to have GEDI returns influenced by vegetation in neighboring fields. Overall, this study has two main goals: (1) to document the performance of GEDI-S2 for crop type mapping around the world, through systematic comparison with existing local reference data, and (2) to identify sources of error in order to guide future efforts to improve performance. Overall, we find that GEDI is a useful resource for advancing the goal of low-cost, timely, and accurate global mapping of crop types.

The following section describes the various datasets used in the study, including any initial processing steps for the data. Section 3 then describes the methods used to map crop height class and evaluate the predictions. Sections 4 presents the main results, while section 5 discusses various sources of errors and potential future directions for improvement. Finally, section 6 briefly summarizes the main conclusions.

\section{Datasets}
This study utilized five main data sources: a global cropland mask used to define cropland areas, GEDI shot returns for cropland areas, Sentinel-2 optical imagery, reference data on crop types from three regions used to train the GEDI classification model, and reference data on crop types from throughout the worlds used to evaluate the performance of our GEDI-S2 tall/short crop predictions. Below we describe each of these, as well as supplementary datasets used to analyze and interpret our results, including a global map that defines the number of growing seasons in each location, and reference crop type maps used to analyze the relationship between peak crop biomass and model errors.

\subsection{Crop mask} \label{data:cropmask}
To identify cropped areas we used the European Space Agency (ESA) \cite{zanaga2021esa} and ESRI \cite{karra2021global} global Sentinel-based 10 m global land cover maps available in the Google Earth Engine (GEE) \cite{gorelick2017google} official and community data catalogs, respectively \cite{awesomegeedatasets}. Both the ESA WorldCover 2020 product and ESRI 2020 Global Land Use Land Cover provide a global land cover map for 2020 at 10 m resolution, the former based on Sentinel-1 and Sentinel-2 data, and the latter based on Sentinel-2 alone. We primarily used the ESA mask, which based on visual inspection better captured cropland in most areas where the two maps disagreed. However, for Kenya and Uganda the ESA mask tended to greatly underestimate cropland area, and to better capture cropland for these countries we therefore merged the two masks, defining a pixel as cropland if either of the two classified it as cropland.

\subsection{GEDI data}
GEDI is a sensor onboard the International Space Station (ISS) that acquires lidar waveforms between 51.6° N and 51.6° S to observe the Earth's surface in 3D.
It is the first spaceborne lidar instrument specifically optimized to measure vegetation structure \cite{dubayah2020global}.
It contains three lasers emitting near-infrared (1064 nm) light. Two of the lasers are full power lasers, with the other coverage laser split into two beams, producing a total of four beams. Each beam is then optically dithered across-track resulting in eight ground tracks (four full power and four cover tracks) spaced 600 m on the ground. 
Shots have an average footprint of 25 m in diameter and are separated 60 m along-track. 

\begin{figure}
	\centering
	\includegraphics[width=1.0\linewidth]{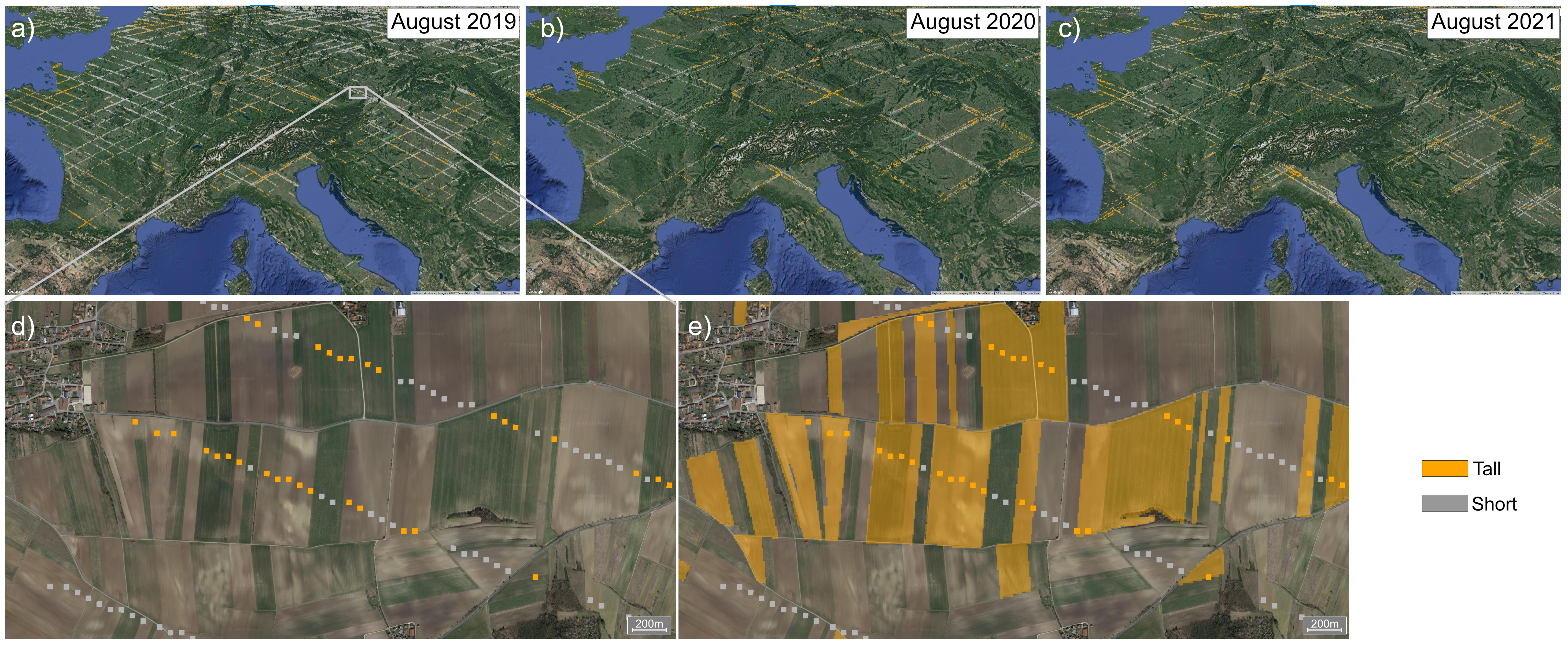}
    \caption{
Spatial locations of GEDI shots over croplands in Europe. (a-c) Show spatial distribution for 2019, 2020, and 2021, respectively. Starting in 2020 higher ISS altitudes caused the clustering of GEDI observations along its orbital track (i.e., "resonance") leaving bigger gaps across tracks.  (d) Zoom in of GEDI shots for 2019 over fields in Austria, color coded by GEDI model predicted class, gray for short, orange for tall crop. (e) Same as (d) but also showing the fields growing maize according to ground truth from Austria parcels dataset, illustrating that GEDI predicted class agrees well with the ground truth.
    }
    \label{fig:GEDIorbits}
\end{figure}

GEDI spatial coverage changes in time. In particular in early 2020 the ISS lifted its orbit, causing GEDI to have “orbital resonance” which means it goes over the same tracks repeatedly while leaving big gaps in between (Figure \ref{fig:GEDIorbits} panels (a)-(c)).

Another important aspect of GEDI is that while its viewing angle is typically near-nadir, it can be rotated by up to 6°, allowing the lasers to be pointed up to 40 km on either side of the ISS ground track. This capability is used to sample the Earth’s land surface as completely as possible, but can also complicate interpretation of the GEDI returns \cite{fayad2020analysis}.

For this study we used the GEDI dataset Level 2A (L2A) and Level 2B (L2B) from April 2019 to December 2021, available in GEE data catalog.
Level 2 data provide information about the vertical distribution of the canopy retrieved from the waveform return at footprint level. The main GEDI product used is GEDI's L2A Geolocated Elevation and Height Metrics Product, which is primarily composed of Relative Height (RH) metrics, which collectively describe the waveform collected by GEDI.
Relative Height (RH) metrics give the height at which a certain percentile of energy is returned relative to the ground. RH are reported at 1\% intervals, resulting in 101 metrics.
The GEDI L2A dataset (\texttt{LARSE/GEDI/GEDI02\_A\_002\_MONTHLY})
is a rasterized version of the original GEDI product , with each GEDI shot footprint represented by a 25 m pixel \cite{healey2020highly}. This rasterization process can introduce an additional geolocation error to the initial GEDI shot error.
The raster images are organized as monthly composites of individual orbits in the corresponding month. RH values and their associated quality flags and metadata are preserved as raster bands.

A secondary dataset L2B was used to retrieved the GEDI view angle (i.e. local beam elevation property). This is available in GEE as table of points (\texttt{LARSE/GEDI/GEDI02\_B\_002}) with a spatial resolution (average footprint) of 25 meters.
At the time of writing the raster version of the L2B dataset (\texttt{LARSE/GEDI/GEDI02\_B\_002\_MONTHLY}) is only partially ingested in GEE, and we therefore used the table. 

\subsection{Sentinel-2}

We used S2 surface reflectance data (Level-2A) present in GEE and filtered out clouds using the S2 Cloud Probability dataset provided by SentinelHub in GEE.
The Sentinel-2A/B satellites acquire images with a spatial resolution of 10 m (Blue, Green, Red, and NIR bands) and 20 m (Red Edge 1, Red Edge 2, Red Edge 3, Red Edge 4, SWIR1, and SWIR2 bands), and together they provide images at a 5-day interval.

To capture crop phenology we extracted S2 imagery for 2019-2021 from January 1 to December 31 for the northern hemisphere, and from July 1 of one year to June 30 of the next for the southern hemisphere. Features were extracted from S2 time series by fitting harmonic regressions to all cloud-free observations in cropped areas. For each spectral band or vegetation index $f(t)$, the harmonic regression takes the form

\begin{equation*}
f(t) = c + \sum_{k=1}^{n} \left[ a_{k} \cos (2 \pi \omega k t) + b_{k} \sin(2 \pi \omega k t) \right]
\end{equation*}

where $a_{k}$ are cosine coefficients, $b_{k}$ are sine coefficients, and $c$ is the intercept term. The independent variable $t$ represents the time an image is taken within a year expressed as a fraction between 0 and 1. The number of harmonic terms $n$ and the periodicity of the harmonic basis controlled by $\omega$ are hyperparameters of the regression.

To determine the harmonic order and omega we sampled multiple locations around the world and compared the harmonic fitting of the time series by varying the hyperparameters. We found that third order harmonics ($n=3$) with $\omega=1$ were a good fit for both regions with one or multiple growing seasons.

We computed harmonic coefficients for four bands and one vegetation index: NIR, SWIR1, SWIR2, RDED4 and GCVI. GCVI is the green chlorophyll vegetation index \cite{gitelson2005remote} computed as
\begin{equation*}
\text{GCVI} = \text{NIR} / \text{Green} - 1
\end{equation*}
This yields seven features per band, for a total of 35 coefficients.
Previous crop type classification studies \cite{wang2019crop,song2021evaluation} have reported the efficacy of using these four bands and VI, demonstrating performance comparable to classification models using all optical bands and a variety of other VIs.

\subsection{GEDI model training dataset} \label{data:GEDI training dataset}

To train the GEDI model to distinguish tall from short crops we used high accuracy crop type labels from 2019 from the three regions used in prior work \cite{di2021combining}: Jilin province in China, Grand Est region in France and Iowa state in USA. 
These regions are major agricultural production areas containing a mix of tall and short crops and have accurate, field-scale crop type maps that are publicly available.
Differences in agricultural practices across regions for same classes could translate in differences in the GEDI waveforms helping the GEDI model to be more flexible and adaptable in different regions as well.

For Jilin, China we used the 2019 crop type map \cite{you202110m}. It maps three major crops in the area (maize, soybean, and rice) at 10 m with an accuracy of 87\%, and F1-scores of 85\% for maize. For Grand Est, France we used the Registre Parcellaire Graphique (RPG) 2019 dataset downloaded from \url{https://www.data.gouv.fr/}. It is a public georeferenced vector product derived via survey. For Iowa, USA we used the U.S. Department of Agriculture 2019 Cropland Data Layer (CDL) at 30 m resolution available in GEE \cite{boryan2011monitoring}. It has an overall accuracy of 90\% and precision and recall for maize exceeds 95\%.

\subsection{Evaluation datasets}
\begin{figure}
	\centering
	\includegraphics[width=1.0\linewidth]{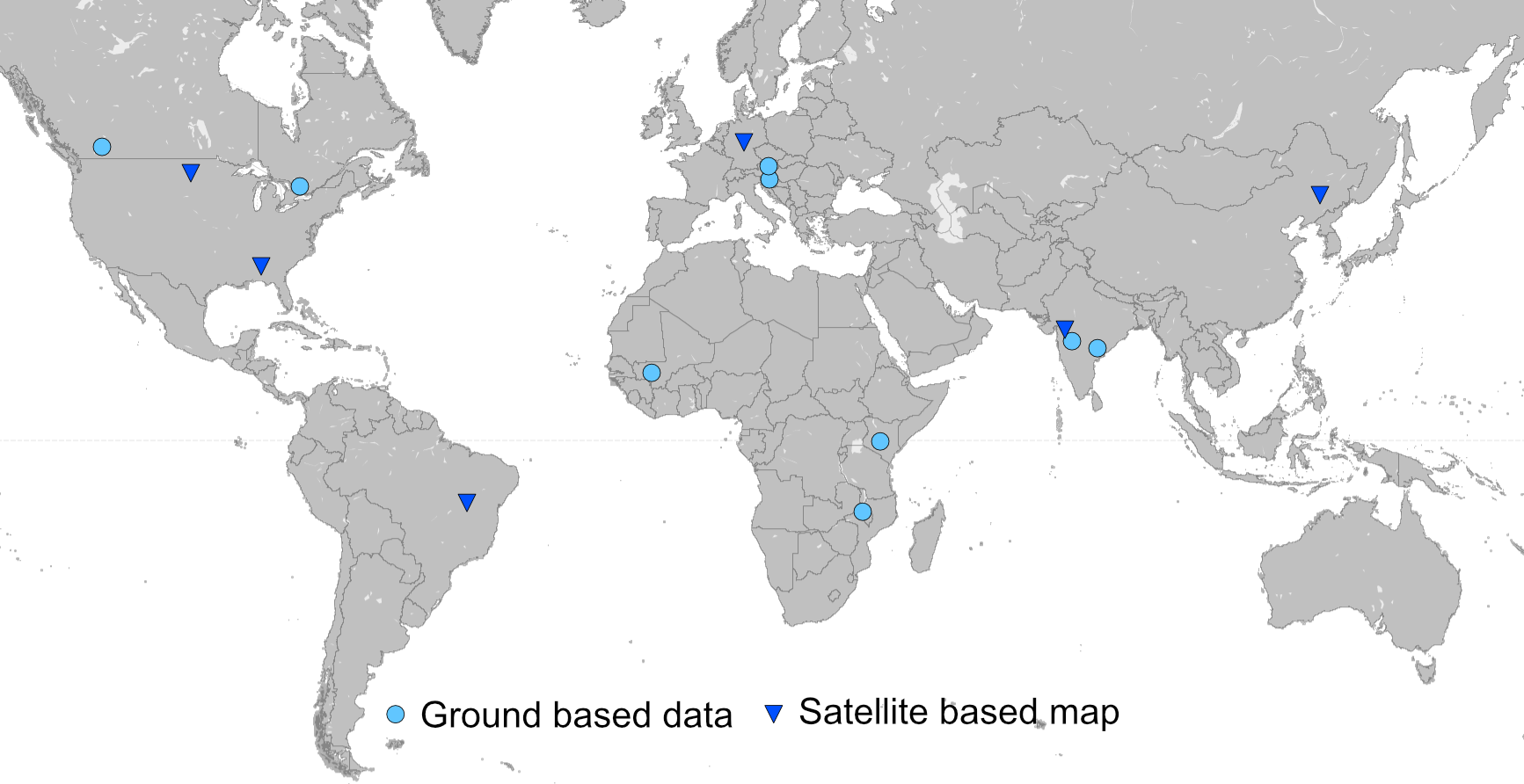}
    \caption{
Distribution of data used for independent evaluation. Ground-based refers to either point or polygon data collected on the ground. Satellite-based refers to maps typically created by combining ground data with remote sensing datasets, such as the Cropland Data Layer in the United States.
    }
    \label{fig:validationLocations}
\end{figure}

To evaluate our product we sought high quality crop type datasets for a diverse set of cropping systems and regions for the 2019-2021 period. We used a combination of field data and crop type maps that were produced by combining field and satellite data. In regions with multiple growing seasons we filtered for crop type labels matching the growing season that the GEDI model predicted for. A map of location and type of reference data is shown in Figure \ref{fig:validationLocations} and a summary of data characteristics including sample size are given in Table \ref{table:validationTable} .

\begin{landscape}
\begin{table*}
\caption{
Characteristics of the regional datasets used to evaluate GEDI-S2 and summary statistics of performance. For each dataset we report the year, the type of reference data, the number of samples and the percentage of reference data labelled as a tall crop. S2-Local model and GEDI-S2 model performance are reported in terms of accuracy, precision, recall, F1 and Kappa scores. Precision, recall and F1 scores are presented as (short, tall) pairs. Canada BC and ON stand for British Columbia and Ontario respectively.
US ND and AL stand for North Dakota and Alabama.
Brazil BA for Bahia.
India UBB for Upper Bhima Basin, TG for Telangana, and MH for Maharashtra.
}

\tiny
\renewcommand{\arraystretch}{1.8}
\centering
\rowcolors{3}{gray!25}{white} 
\begin{tabular}{*{1}{l}*{4}{c}|*{5}{c}|*{5}{c}}\hline
\multirow{2}{*}{\textbf{Region}}  & \multirow{2}{*}{\textbf{Year}}  & \multirow{2}{*}{\textbf{Type}}  & \multirow{2}{*}{\textbf{Samples}}  & \multicolumn{1}{c}{\textbf{Tall}}  & \multicolumn{5}{c}{\textbf{S2 Local}} & \multicolumn{5}{c}{\textbf{GEDI-S2}}  \\
 &  &  &  & \% & Accuracy & F1 & Precision & Recall & K-score & Accuracy & F1 & Precision & Recall & K-score \\
\hline
\csvreader[late after line=\\, late after last line=\\\hline]
{tables/GEDI_Validation2.csv}{1=\pa,2=\pb, 3=\pc, 4=\pd, 5=\pe, 6=\pf, 7=\pg, 8=\ph, 9=\pi, 10=\pl, 11=\pm,12=\pn,13=\po,14=\pp,15=\pq}{\pa  & \pb  & \pc & \pd & \pe & \pf & \pg &\ph &\pi &\pl &\pm &\pn  &\po &\pp &\pq}
\end{tabular}

\label{table:validationTable}
\end{table*}

\end{landscape}


\subsubsection{Ground-based reference data}

\paragraph{Europe} 
\citeasnoun{schneider2021eurocrops} contains harmonized agricultural parcels information data from regions in Austria (2019), Denmark (2020) and Slovenia (2019). The parcel data are based on publicly available self-declared crop reporting datasets, gathered for the purposes of subsidy payments. We focused on Austria and Slovenia, since the Denmark dataset is outside the GEDI spatial coverage. Major cover types include maize, wheat, barley, and pasture, with parcels indicating only one crop per year.

\paragraph{Canada}
\citeasnoun{aafcground} is a collection of thousands of points identifying crops types and occasionally other land cover types across Canada from 2011 to 2021.
These point sources are used by Agriculture and Agri-Food Canada (AAFC) as training or reference sites for the creation of the Annual Crop Type map.
Major crops include mixed forage, maize, soybean and alfalfa.


\paragraph{Malawi}
Field boundaries for three crops -- groundnuts, maize, and soybean -- were collected in five districts of Malawi (Lilongwe, Ntchisi, Kasungu, Salima, and Mzimba) for 2021 as part of research on the groundnut value chain conducted by the AgroHitech Innovation and Advisory Consortium for the Peanut Innovation Lab. 

\paragraph{Kenya}
The Global Agriculture Monitoring initiative of the Group on Earth Observation, called Copernicus4GEOGLAM, collected ground reference data during field surveys in three countries -- Uganda, Tanzania and Kenya \cite{geoglam} -- for both the 2021 long rain and 2021-22 short rain seasons. The georeferenced ground data were used by Copernicus4GEOGLAM to train random forest models to map crop type using S2 imagery as input. Given the fairly low accuracies of the resulting maps (e.g., maize F1 scores were often below 0.6), we utilized only the field data for our evaluation. We focused on the Kenya point dataset for the 2021-2022 short rain season, which had the greatest number of points overlapping with the season of our GEDI-S2 predictions. Main crops in this region include maize, tea, sugarcane and potatoes.


\paragraph{India, Plantix} 
Plantix is a free Android application created by Progressive Environmental and Agricultural Technologies (PEAT).
The Plantix app is used by farmers who submit photos of their crops seeking help to diagnose and treat crop diseases.
As part of the disease diagnosis, PEAT uses a convolutional neural network to assign crop labels based on the submitted photos. We used these data in the Indian states of Maharashtra and Telangana, where the accuracy of Plantix crop type labels exceeds 0.90 for most major crops. 
These data have been cleaned to remove location inaccuracy (keeping only submissions with GPS accuracy better than 10 m), as suggested by previous work by \cite{wang2020mapping}.
To match the timing of the GEDI-S2 predictions, we filtered the Plantix data for the 2021 kharif season based on photo submission timing. Major crops in these states include cotton, maize, rice, sugarcane and peanut.

\subsubsection{Satellite-based reference data}

\paragraph{United States}
The Cropland Data Layer (CDL) produced by the United States Department of Agriculture (USDA) provides yearly crop type maps across the conterminous US at 30 m spatial resolution \cite{boryan2011monitoring}.
Maps are based on Landsat and other satellite imagery using training data from the Farm Service Agency (FSA). For validation we chose two states, North Dakota and Alabama, that were far from the conditions and locations of the Iowa locations used in the training data. Accuracy of CDL on FSA labels are available in the CDL metadata, with precision and recall for maize for 2019-2021 higher than 81\% and 85\% in North Dakota and Alabama, respectively.

\paragraph{Germany} 
National scale crop type maps for Germany were recently produced for 2017-2019 \cite{blickensdorfer_lukas_2021_5153047} and  2020 \cite{schwieder_marcel_2022_6451848}. These maps are generated using a random forest classifier based on Sentinel-1, Sentinel-2 and Landsat time series, with parcel data used for training. More details about the underlying data and methods can be found in \citeasnoun{blickensdorfer2022mapping}.
Overall accuracy for 2019 is 78\%, with precision and recall for the maize class of 90\% and 83\%, respectively. Major crops include maize, wheat and barley.

\paragraph{Brazil}
Annual soybean maps were recently produced for South America at 30 m resolution between 2000 and 2020 by combining Landsat and MODIS satellite observations and sample field data \cite{song2021massive}. These maps are available in GEE as (\texttt{projects/glad/soy\_annual\_SA}).
For evaluation, we focused on western Bahia in 2020, since this region grows maize and soy in the main season and has only one primary growing season per year. We evaluated only recall on soy, since other short crop like cotton are also grown in the same season but are not distinguished from other non-soy crops in their study. 
Accuracy for 2020 is not reported, but for the years 2017-2019 they report overall accuracies of 96\%, 94\% and 96\%, respectively, with high and balanced producer’s and user’s accuracies. 

\paragraph{China}
For China, we used the same 2019 crop type map described by \citeasnoun{you202110m} in section \ref{data:GEDI training dataset} used in training the GEDI model. For validation, we used a random sampling over the four Northeast regions (Liaoning, Nei Mongol, Jilin and Heilongjiang), which span a much larger area than used in the training sample from Jilin. 

\paragraph{India (Lee et al. 2022)}
\citeasnoun{lee2022mapping} produced a map of sugarcane area in the Upper Bhima Basin, a major sugarcane producing region in Maharashtra, India. Their 10 m resolution map is based on crowdsourced Plantix data and a neural network applied to S2 data.
Reported overall accuracy for sugarcane vs. not sugarcane was 77\% (85\% precision and 67\% recall). Major crops in the region include sugarcane, cotton and rice.

\subsection{Global 10 km gridded data} 
MapSPAM 2010 v2.0 is a global spatially explicit dataset on agricultural production created by the International Food Policy Research Institute (IFPRI) \cite{DVN/PRFF8V_2019}.
The Spatial Production Allocation Model (SPAM) provides estimates of crop distribution for 42 crops at 5 arc-minute resolution (around 10 km resolution at the equator) relative to the 2009-2011 time period.
For  comparison with our tall crop map, we focused on four tall crops: maize, sugarcane, sunflower and cassava (using the layer corresponding to physical area -- all technologies together). SPAM is based on administrative data and does not report accuracies at the grid cell level. In addition, it refers to a time period outside of our GEDI study period of 2019-2021. These caveats limit the usefulness of comparing SPAM to our predictions; however, we include it as a useful reference point given that SPAM is widely used by the research community.

\subsection{Number of growing seasons per year} 
The Anomaly hostspots of Agricultural Production (ASAP) system is an online decision support tool for early warning about production anomalies developed by the Joint Research Center (JRC) of the European Commission.
ASAP has produced several maps including satellite-based phenology information, which are computed from the long-term average of MODIS NDVI data at $0.01\degree$ resolution \cite{rembold2019asap}. We downloaded the phenology layer that defines the number of growing seasons (1 or 2 seasons) \cite{ngws}, and aggregated this information at $5\degree$ based on the majority of the crop pixels' seasonality.

\subsection{Digital elevation model (DEM)} 
The Shuttle Radar Topography Mission SRTM V3 (SRTM Plus) \cite{farr2007shuttle} digital elevation data product is provided by NASA JPL at a resolution of 1 arc-second (approximately 30 m) and is available in GEE. 
We calculated the slope in degrees from the terrain DEM in GEE.

\subsection{Reference maps for error analysis}
As described below, we hypothesize that errors in our GEDI-S2 predictions were often related to low biomass of the tall crop. To further investigate this, we utilized two additional crop type maps that provided wall-to-wall coverage in countries where our preferred data for evaluation covered only a subset of fields. Widespread coverage was needed to ensure a wide range of biomass values for pixels in the reference map that overlap with the GEDI shot locations.

\subsubsection{Canada}
The Earth Observation Team of the Science and Technology Branch at Agriculture and Agri-Food Canada (AAFC) have created Annual Crop Inventory maps that are accessible in GEE. These maps are generated using a combination of crop type labels from crop insurance data and ground-truth information collected across the country to train a decision tree model based on optical and radar satellite images. Maps have a spatial resolution of 30 m and an accuracy of at least 85\%.

\subsubsection{Kenya}
As described above, in addition to field data, the Copernicus4GEOGLAM produces end-of-season crop type maps for each country and season where field data was performed \cite{geoglam}. In our error analysis, we used the long rains map for Kenya, which possesses the highest F1 score for maize among the various countries and seasons. For this map, overall crop type accuracy is 80\%, and F1 for maize is 0.64.

\section{Methods}
\label{sec:methods}

Here we describe the steps taken to create and evaluate wall-to-wall maps of crop type height, using a combination of GEDI and S2 as input. The sections below describe in detail each of the six steps in this process:

\begin{enumerate}
    \item Train a model, called \textit{GEDI model}, that uses GEDI features to classify locations as having short crops, tall crops, or trees;
    \item Apply the GEDI model to GEDI shots acquired from cropland areas globally for three years of 2019-2021;
    \item Tile the globe into $5\degree \times 5\degree$ grid cells;
    \item Determine the optimal month to predict tall crops for each grid-cell;
    \item Train a local \textit{GEDI-S2 model} for each grid-cell based on GEDI predictions in the 3-month time window around the optimal month;
    \item Evaluate results against local reference data and other global maps.
\end{enumerate}

\subsection{GEDI model training}
Following \citeasnoun{di2021combining}, we began by defining a random forest model to classify GEDI shots in 3 crop height classes: short, tall or tree. To train the model we used labels from three areas with high-quality crop type maps for 2019: Jilin in China, Grand Est in France, and Iowa in the United States (see section \ref{data:GEDI training dataset}). Crop type labels were sampled at GEDI shot locations and assigned a \textit{tall} label for maize class and a \textit{short} label to remaining short crops. We also defined a third \textit{tree} class, for shots with RH100 greater than 10 m.

We used all GEDI shots in August 2019 over the 3 regions since our previous study \cite{di2021combining} showed August to be a good time to distinguish maize from other short crops in these regions. To reduce the number of features, since consecutive RH metrics are highly correlated with each other, we sampled a metric every 5\%  and omitted RH in the middle of the RH profile based on feature importance analysis. In total, 11 RH metrics were used: RH0, RH5, RH10, RH15, RH20, RH25, RH30, RH85, RH90, RH95, and RH100. 

The features and labels were used in a random forest model, implemented in GEE. Data were split into 80\% training and 20\% test points. To minimize spatial correlation across the training and test sets, we binned the shots by their lat/lon into $0.5^{\circ} \times 0.5^{\circ}$ bins and GEDI shots in each bin were placed entirely in either the training set or test set. The overall test accuracy across the three regions was 0.885, with F1 scores for short, tall and tree classes of 0.863, 0.898 and 1.000, respectively. 




\subsection{GEDI model predictions}
The random forest model described above was then applied to all GEDI shots in cropland pixels, according to the crop mask described in Section \ref{data:cropmask}. The predicted class was saved along with the prediction probabilities (the fraction of trees in the random forest model that predicted the class) as a measure of confidence.
Figure \ref{fig:GEDIorbits} illustrates these predictions for a selection of GEDI orbits, with shots colored orange for tall and gray for short based on the GEDI model predictions.

The predicted shots were filtered to retain only high quality shots to use as labels in subsequent steps.  First, we removed shots with a quality flag value of zero in the original GEDI returns, which indicates poor quality, as well as shots with a non-zero degrade flag, which indicates poor geolocation. We then removed low confidence predictions (lower than 0.8) to have more confidence in the GEDI-generated labels.

Another step that proved essential was to filter out shots with low view angle and on high slope terrain since both factors can affect the accuracy of the GEDI model predictions. We refer to view angle as the angle between the off-nadir beam and the ground.
Prior work has revealed that small changes in view angle can increase errors for models based on GEDI returns \cite{fayad2020analysis,fayad2022comparative}. In particular, existing analysis recommends removing observations where the view angle was below 1.5 rad, or roughly $86\degree$ \cite{fayad2022comparative}. 

\begin{figure}
	\centering
	\includegraphics[width=1.0\linewidth]{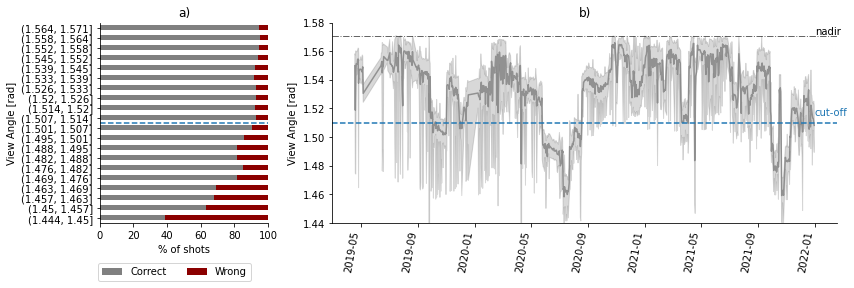}
    \caption{
(a) Effects of GEDI view angle on the accuracy of the GEDI model predictions in the Central United States. (b) The temporal variation of view angle over the study period, with shading showing the min-max values. GEDI can be rotated up to 6° (0.1 rad) from nadir. In some periods, such as summer 2020, GEDI was systematically targeting reference ground tracks that were further off nadir than other times. Dashed line labeled ``cut-off'' indicates the threshold used to filter GEDI shots in this study.}
    \label{fig:viewangle}
\end{figure}

To explore the appropriate threshold for our application, we considered shots for the US Corn Belt where we have confidence in the reference data from CDL, and where the view angle property is available in GEE at the GEDI shot level. The GEDI model prediction errors (treating CDL as truth) were evaluated for different levels of view angle, as shown in Figure \ref{fig:viewangle}(a). At low view angles, errors are as high as 60\%. Above the recommended threshold of 1.5 rad, however, errors are below 10\% and fairly insensitive to additional increases in view angle. We therefore adopted a threshold of 1.5 rad for further analysis.

The GEDI view angle varies over time as shown in Figure \ref{fig:viewangle}(b). View angles were particularly low in June and July of 2020, causing the removal of most shots during the peak of the growing season in many regions. Other periods of frequent observations with low view angles include late 2019 and late 2021. Unfortunately, at the time of writing, information on the GEDI view angle was not yet available at the shot level for all shots globally in the GEE catalog. To create a view angle filter, we sampled orbits over a longitudinal transect and aggregated these data, averaging the view angle for each beam on each day. We then removed all shots from beams and days with an average view angle above 1.5 rad. Although this was a pragmatic way to filter out data with low view angle globally, future versions would likely benefit from accounting for view angle at the shot level, to account for variation by latitude and over time within the day.

We also removed shots on high slope terrain, defined as areas with slope higher than $5\degree$. GEDI metrics are dependent on topographic slope \cite{fayad2021terrain}, and given the relatively small height signal being used by our model to classify tall vs. short crops, the effect of topographic slope are potentially important. Based on analysis of CDL in the United States, similar to the view angle analysis presented in Figure \ref{fig:viewangle}(a), a slope below $5\degree$  was deemed sufficient to avoid artifacts from the terrain. As cropland is typically situated on flat or nearly-flat land, this filter removed only a small fraction of GEDI shots.  

\subsection{Model grids}
The filtered GEDI model predictions provide labels with which to train a model that takes S2 data as input. However, we did not expect a single model to be applicable globally, since the timing of growing season and mix of crops differs across the world. Building on prior approaches \cite{healey2020highly,potapov2021mapping} we instead sought to develop locally-calibrated models. We defined a grid within the GEDI coverage (between 51.6\degree N and 51.6\degree S) with $5\degree \times 5\degree$ cells. Although more localized models would potentially improve performance in some regions and years, the choice of grid cell size was dictated by the orbital resonance of GEDI in 2020 and 2021. That is, a finer grid would often have cells that have very few GEDI observations because of the large gaps in GEDI coverage in those years. 

To reduce computation, we only processed grid cells for which more than 5\% of S2 pixels were classified as cropland, yielding 238 cells. The grid cells that we kept cover an area that comprises 93\% of the total crop area within the latitude bands of GEDI coverage. 

\subsection{Optimal timing}
For each grid cell, we defined the optimal month to classify tall vs. short crops as the month in which the highest percentage of GEDI shots were predicted as tall. Specifically, we combined the 3 years of GEDI model predictions by month, computed the percentages of tall and short shots by grid cell and month, and then selected the month for each cell when the percentage of tall shots was highest.  

\subsection{GEDI-S2 models}
For each grid cell and for each year, we separately trained a local 2-class (tall vs. short) S2 model using the GEDI predictions for the relevant time as labels and harmonic coefficients as features. We refer to these as GEDI-S2 models, with a unique model for each grid cell and year. To account for variations in the timing of the growing season within the $5\degree$ grid cells, we considered a three-month window centered on the optimal month.
We created GEDI-S2 predictions for individual months and then combined the predictions on a pixel basis, with pixels classified as tall if the predicted class was tall in any of the three months. 

The result of this process was a wall-to-wall 10 m resolution map of tall and short crops for all cropland pixels in the $5\degree$ grid cells. 
To reduce computation, we only applied the GEDI-S2 models to grid cells where the percentage of tall shots is higher than 4\%, i.e. 201 grid cells per year. 
Since GEDI data was not always available in all the regions in the time window of interest, the number of grid cells processed is 1562, less than the expected 1809 = $201 (grid cells) \times 3 (months) \times 3 (years)$.
This resulted into 189 unique locations in 2019, 457 in 2020 and 201 in 2021, for a total of 590 grid cells for the 3 years.

For this local training, we omitted all shots where the GEDI model predicted the tree class, as these were viewed as likely to be a mixture of crops and trees within the GEDI footprint, which at 25 m diameter is more than four times larger than the 10 m S2 pixel. Thus, predictions of tree were viewed as unreliable labels for a 2-class model focused on S2 pixels classified as cropland. 

To minimize spatial artifacts when mosaicking adjacent cells, we created predictions for pixels in a $0.5\degree$ buffer around each cell and mosaicked the overlapping predictions taking the predictions in the cell with higher GEDI-S2 accuracy.

\subsection{Evaluation of GEDI-S2 predictions}

The first evaluation of GEDI-S2 predictions is against reference data from around the globe (Table \ref{table:validationTable}). All reference data were ingested in GEE for comparison with GEDI-S2 predictions. For regions with ground-based point or polygon data, we used all fields for evaluation. In the case of polygons, the centroid of the polygon was used to define the relevant pixel from the GEDI-S2 predictions for comparison. For regions where crop type maps were used, we randomly sampled the maps using 2,000 to 4,000 points and removed the ones without a specific crop type label to create a reference dataset.

Because some datasets contain as many as 100 crop types, for each reference dataset we selected the 10 most common crops for evaluation, which typically represents more than 90\% of the crop areas in the reference regions evaluated. From specific crop type labels, we generated a binary tall/short classification, with maize and sugarcane defined as tall and all other crops defined as short (none of our evaluation data had sunflower or cassava among the 10 most common crops). For each evaluation dataset, we report the accuracy, precision, recall, F1 and Kappa scores, using the following equations \cite{grandini2020metrics}:

\[ Accuracy
  = \dfrac{TP+TN}{TP+TN+FP+FN}
\]

\[ Precision
  = \dfrac{TP}{TP+FP}
\]

\[ Recall
  = \dfrac{TP}{TP+FN}
\]

\[ F1\ score
  = \dfrac{2 \times Precision \times Recall}{Precision+Recall}
\]

\[ Kappa\ score
  = \dfrac{P_{o}-P_{e}}{1-P_{e}}
\]
where
\begin{itemize}
\item {True Positive, $TP$, is the number of samples labelled as positive by the model that are actually positive}
\item {False Positive, $FP$, is the number of samples labelled as positive by the model that are actually negative}
\item {True Negative, $TN$, is the number of samples labelled as negative by the model that are actually negative}
\item {False Negative, $FN$, is the number of samples labelled as negative by the model that are actually positive}

\item {$P_{o}$ is the proportion of observed agreement, i.e. the accuracy achieved by the model}
\item {$P_{e}$ is the proportion of agreements expected by chance}

\end{itemize}

Our second evaluation compares GEDI-S2 predictions against an S2 model trained locally within each reference region. These \textit{S2-Local} models provide benchmarks that represent how well a model trained on local field data would perform. To conduct this analysis, we exported S2 harmonic features at the same reference field locations and trained a local S2 model based on binarized tall/short field labels. We used the \texttt{RandomForestClassfier} implemented in Python's \texttt{scikit-learn} package, 
setting similar hyperparameters to the GEDI-S2 random forest models implemented in GEE. Reference data were binned by their lat/lon into $0.5\degree \times 0.5\degree$ bins, and data in each bin were placed entirely in either the training set or test set using a 80\%/20\% train/test split. We ran the S2 classifier multiple times using each time a different train/test split and reported the average S2-Local model performance metrics.

Our third and final comparison is against the 2009-2011 SPAM dataset. To enable comparison, we aggregated our maps of tall/short crops to the same resolution as the SPAM gridded dataset ($\sim$10 km resolution at the equator). Tall crop area in SPAM was defined as the sum of the area of four crops: maize, sugarcane, sunflower and cassava.


\section{Results}\label{results}
\subsection{GEDI predictions during optimal months}

\begin{figure}
	\centering
	\includegraphics[width=1.0\linewidth]{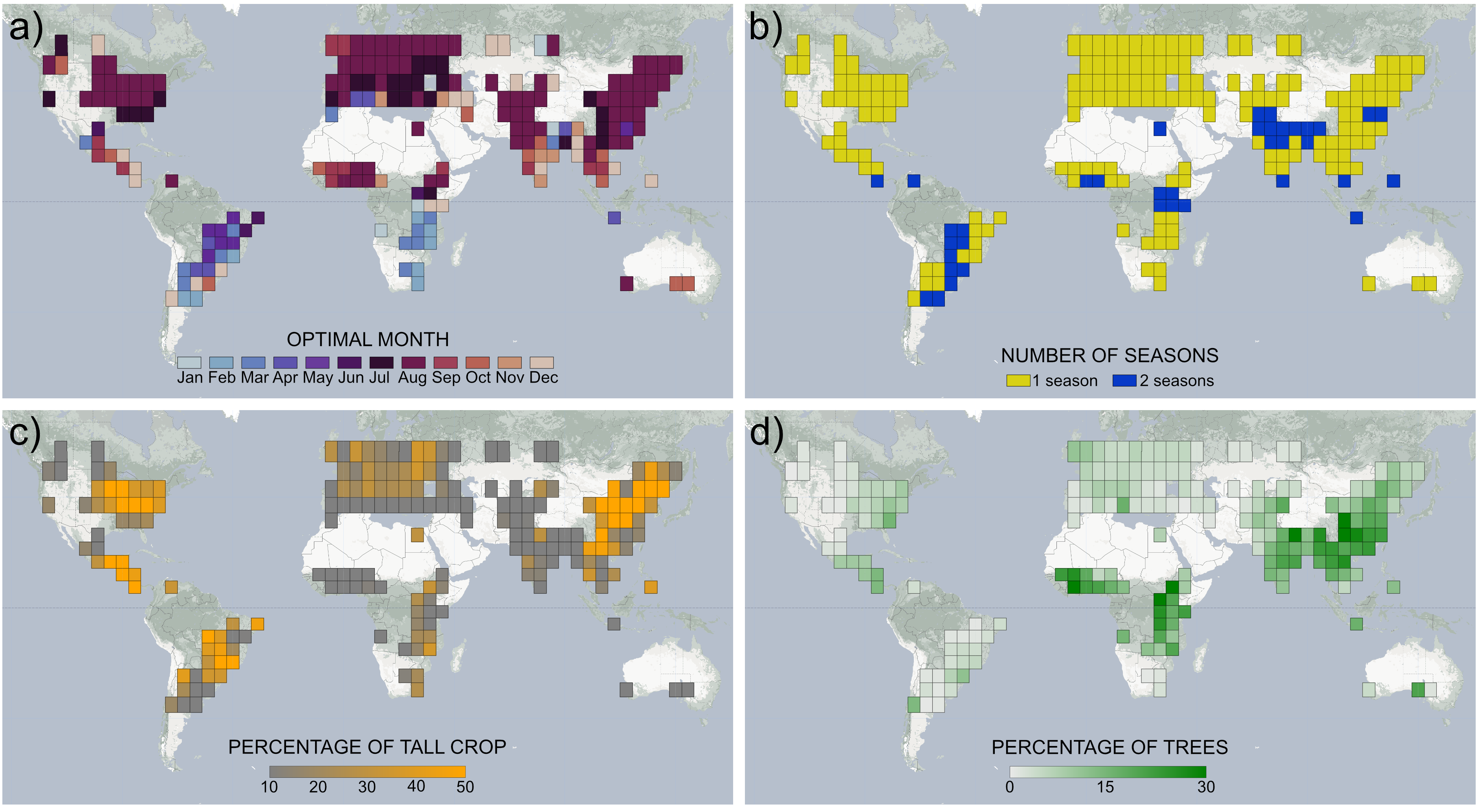}
    \caption{
    Characteristics of 5\degree x 5\degree grid cells where GEDI-S2 was applied. (a) The optimal month to identify tall crops, defined as the month with the greatest proportion of shots classified as tall, (b) the number of growing seasons per year based on the Anomaly hostspots of Agricultural Production (ASAP) phenology information dataset \protect\cite{ngws}, (c) the percentage of shots classified as tall, and (d) the percentage of shots classified as trees.
    }
    \label{fig:optimalmonth}
\end{figure}

The fraction of GEDI shots classified as a tall crop generally reaches its peak during the latter months of the growing season, such as August throughout much of the Northern Hemisphere (Figure \ref{fig:optimalmonth} (a)). This period coincides with the grain filling stage after flowering, when maize has reached its peak height, but before fall months when harvesting begins. We refer to this month of peak tall crop percentage as the optimal month for using GEDI to distinguish tall and short crops. In most regions, the optimal month is stable across space, with most neighboring cells differing by no more than one month. Exceptions to this pattern are evident in cases where tall crops are a very small percentage of the total crop area (e.g. western Canada) or where two or more growing seasons occur throughout the year (e.g., Brazil and India) (Figure \ref{fig:optimalmonth}(b)). In both cases, this can lead to two or more months having very similar tall percentages, making the optimal month less stable. 

The fraction of GEDI shots from croplands classified as tall crops during these peak months exhibits large spatial variation, with a pattern that coincides closely with known areas of maize production (e.g. eastern United States, eastern China, Brazil). The fraction reaches as high as 75\% in eastern China and parts of Central America, where maize dominates the summer growing seasons (Figure \ref{fig:optimalmonth}(c)). Beyond these regions, shots classified as short crops are generally the dominant class, although a sizable fraction of shots in Africa and Asia were classified as trees (i.e. RH100 $>$ 10 m) (Figure \ref{fig:optimalmonth}(d)).
The presence of trees in areas classified as cropland likely reflects both a higher proportion of trees in crop fields in these regions, as well as a lower precision of the ESA cropland map in smallholder systems common in these regions. 

\subsection{GEDI-S2 model training}
When using the GEDI predicted class (i.e. tall vs. short crop) as labels to train a model based on S2 harmonics, we find that the S2 models are generally able to explain a very high fraction of variability in the GEDI class. In 95\% of grid cells (1488 out of 1562 total grid cells), the test accuracy for the model on GEDI shots held out of training was over 0.85. This indicates that tall crops in a region (e.g. maize or sugarcane) are typically distinct enough from other crops in the feature space of the S2 harmonics. In a small number of locations (21 out of 590 cells over the 3 years), the GEDI-S2 training was relatively poor with a test accuracy averaged across all months of less than 0.85. These cells typically occur in regions with high topographic variation, a factor that is known to affect GEDI returns and reduce the accuracy of tree height models based on GEDI \cite{fayad2021terrain}. Fortunately, the small number of locations with poor GEDI-S2 training performance indicates that, for most agricultural settings, topographic variation or other sources of error are not a major impediment to using GEDI to identify tall crops. We emphasize that this statement only applies to GEDI shots that have first been filtered for view angle, as including shots with high view angles leads to substantial degradation of performance.

\subsection{GEDI-S2 model evaluation}
\begin{figure}
	\centering
	\includegraphics[width=1.0\linewidth]{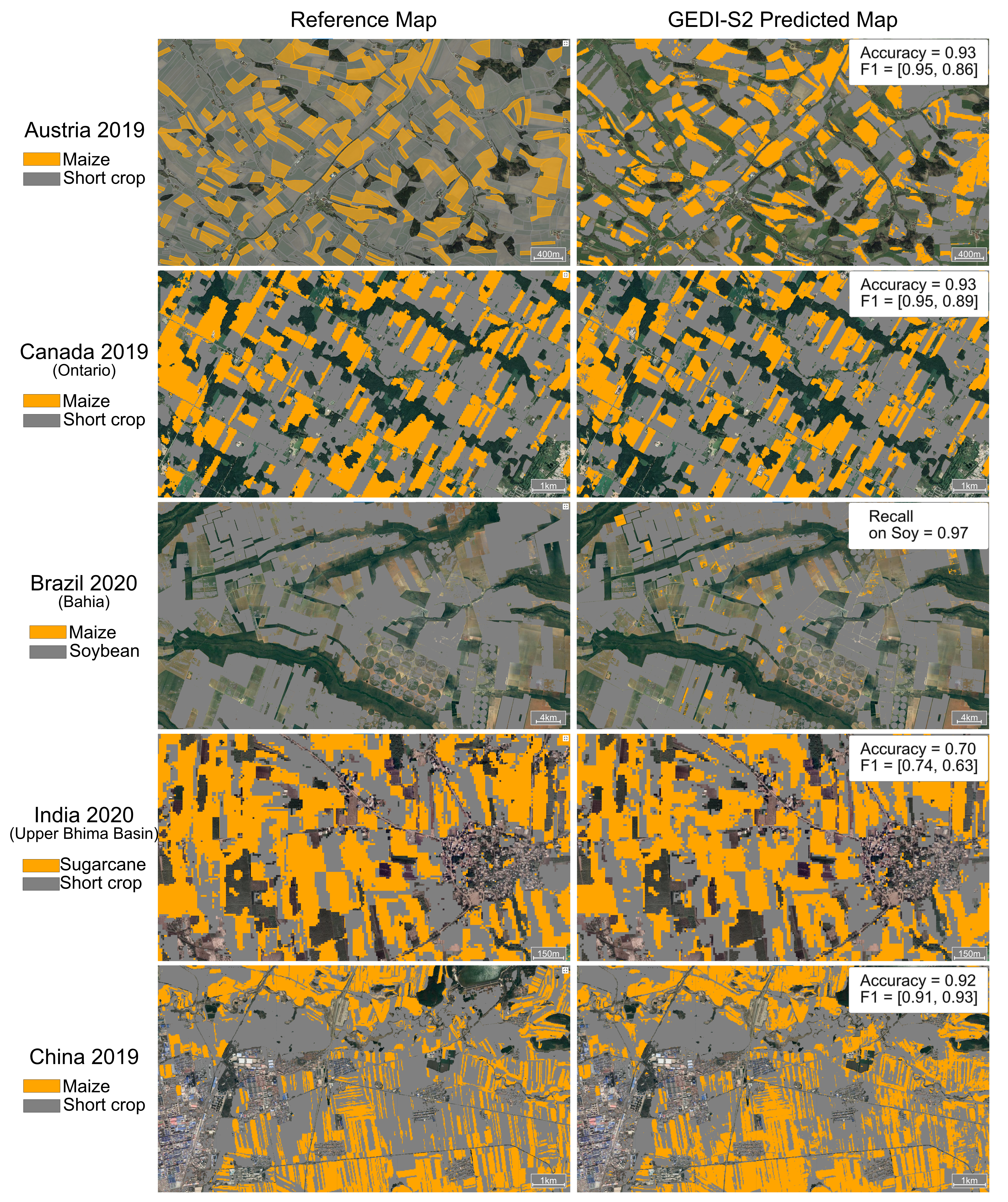}
    \caption{
Visual comparison of reference maps (left) and predicted crop classes from the GEDI-S2 model (right). The accuracy and F1 scores in top-right of each row refer to the entire region, not just the small areas displayed in the figure. F1 scores are presented as [F1-short, F1-tall]. In Brazil, only recall for short crops is reported since the reference map contained only soybean locations.
    }
    \label{fig:validation}
\end{figure}

The predicted crop class from the GEDI-S2 model is able to closely reproduce reference maps of tall crops in many (but not all) cases (Table \ref{table:validationTable}, Figure \ref{fig:validation}). Performance was similar for both ground-based measurements of crop type (either as points or field polygons) and raster maps of crop type developed through a combination of ground and satellite data. In both cases, we compare the GEDI-S2 model performance to the performance of a model trained on the local ground truth with the same S2 features as used in the GEDI-S2 model (S2-Local). This comparison identifies the impact of substituting GEDI measurements for ground observations.

\begin{figure}
	\centering
	\includegraphics[width=1.0\linewidth]{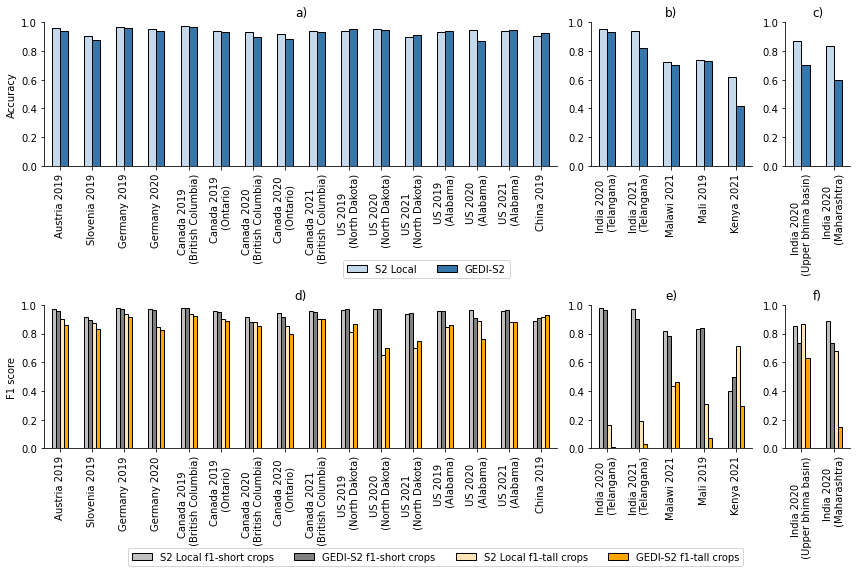}
    \caption{
    Comparison of GEDI-S2 model performance in terms of accuracy (a-c) and F1 scores (d-f) with the S2-Local model (i.e., a Sentinel 2 model trained on the local reference labels). In most regions, the performance of GEDI-S2 was very close to and occasionally even exceeded that of the S2-Local model (a,d). In some areas, both approaches struggled with tall crops (b,e), while in others GEDI-S2 performed worse than the S2-Local model (c,f).
    }
    \label{fig:accuracy}
\end{figure}

Given the large number of validation regions, we discuss the results in terms of clusters of similar behavior rather than discussing each region in detail. The first cluster includes regions where both the local S2 and GEDI-S2 models show high performance, with accuracies typically above 0.9 and F1 scores for both short and tall crops that typically exceed 0.8. Among the regions in this category are North America, Europe, and China (Figure \ref{fig:accuracy}, panel (a) and (d)). Visual inspection of maps generated by training on ground data vs. GEDI corroborate the strong performance of GEDI (Figure \ref{fig:validation}). Performance in Brazil appears similarly strong (Figure \ref{fig:validation}), although because the reference map only identified soybean locations, we cannot calculate total accuracy or F1 scores but only recall for the short crop class. 

In a second set of regions, the performance of S2-Local and GEDI-S2 models were lower, but were similar to each other (Figure \ref{fig:accuracy}, panels (b) and (e)). This situation, which occurs in parts of India and Africa, indicate where both approaches struggle to accurately map crop classes, particularly the tall class. One plausible reason for this is that the phenological and spectral differences between different crops are smaller in these regions, so that harmonics-based features are less informative. More sophisticated features, such as based on convolutional neural networks, could help to improve both models but are beyond the scope of the study. 

A third and final category of performance includes regions where GEDI-S2 performs notably worse than S2-Local models (Figure \ref{fig:accuracy}, panels (c) and (f)). In these cases, which we observe primarily in India, the use of labels from GEDI rather than local ground data incurs a loss of accuracy. Here the problem is unlikely to be either uninformative harmonic features or noisy reference data, both of which would also affect the local S2 model. Instead, our GEDI model appears to be mislabeling many points, with a tendency in particular to overstate the percentage of short crops. We further analyze the sources of these errors in the discussion section (\ref{errors}). 

\subsection{The global distribution of tall crops}
\begin{figure}
	\centering
	\includegraphics[width=1.0\linewidth]{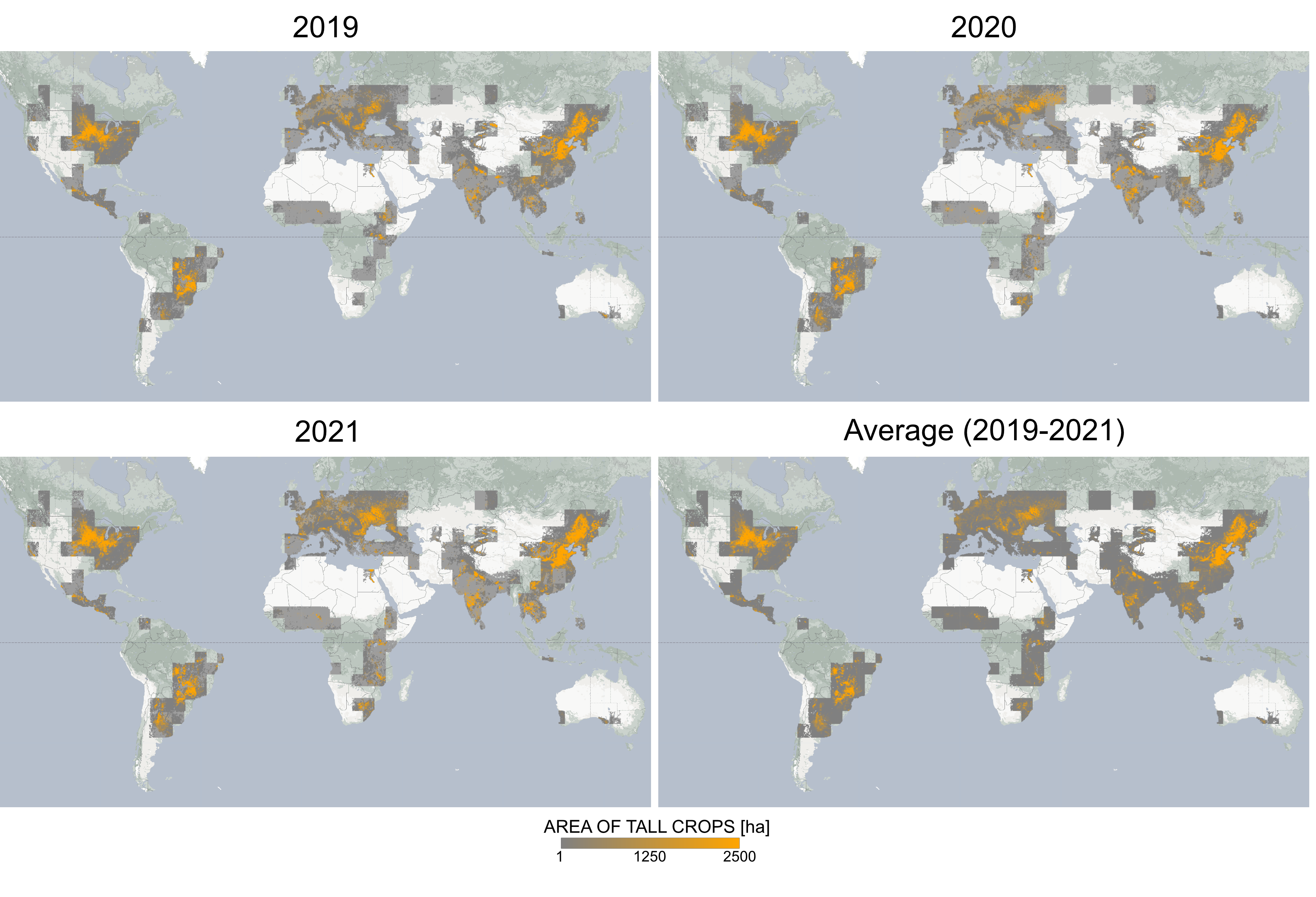}
    \caption{
GEDI-S2 global maps gridded at 10 km resolution. 
For the individual years, cropped areas with peak GCVI above 4 are mapped using the color scale shown, while those with peak GCVI below 4 are mapped with a lighter shade of gray. Low peak GCVI is used as an indicator of where GEDI-S2 is prone to underestimating tall crop area.
    }
    \label{fig:maps3y}
\end{figure}

The overall pattern of tall crop area estimated by our model is shown for each year in Figure \ref{fig:maps3y}. In general, this map corresponds to known areas of maize production, as expected because maize is the most widespread tall crop in the world (Figure \ref{fig:faostats}). However, it also indicates areas with significant sugarcane area (e.g., western Uttar Pradesh in India, the eastern coast of South Africa, and the Philippines) as well as sunflower area (e.g., Romania and Ukraine). 

\begin{figure}
	\centering
	\includegraphics[width=1.0\linewidth]{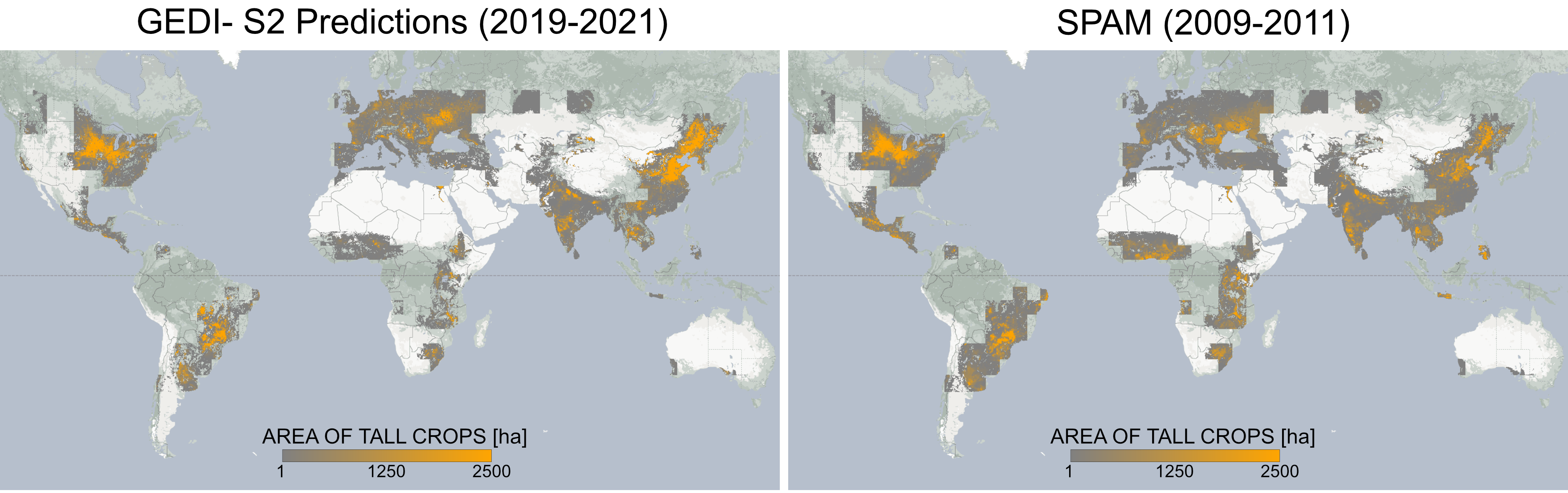}
    \caption{
Comparison of 3-year average GEDI-S2 predictions and the SPAM map for locations where both data are available, with years indicating the respective time periods for each dataset. GEDI-S2 predictions have been gridded at 10 km to match SPAM spatial resolution. Tall crop area in the SPAM map represents the total area of maize, sunflower, sugarcane and cassava.}
    \label{fig:GEDIvsSPAM}
\end{figure}

Our maps agree qualitatively with MapSPAM, a commonly used gridded dataset of crop area at 10 km resolution (Figure \ref{fig:GEDIvsSPAM}). Specifically, our total tall crop area correlates spatially with the sum of maize, sugarcane, and sunflower area from MapSPAM. The agreement between the two is not expected to be exact, for at least two reasons. First, whereas we only consider crop area for a single growing season in each region, MapSPAM includes all annual harvested area. Second, MapSPAM corresponds to an older time period of 2009-11, which is ten years earlier than our estimates of 2019-2021. Despite these differences, we see broad similarities between the two maps, with spatial correlations within several countries exceeding 0.4 (e.g. United States, Canada, China, Brazil, South Africa). Some countries exhibit much greater tall crop area in our maps than in MapSPAM, which correspond to areas with known rapid expansion of maize area since 2010. For example, according to FAO statistics, the maize area in China increased by roughly 30\% from 2010 to 2020, and in Argentina maize area more than doubled over this time period. Thus, the higher estimates in our maps are consistent with expectations on where maize area has been expanding most rapidly.

The broad agreement with MapSPAM at 10 km is an additional confirmation of the GEDI-S2 approach, beyond the pixel-level evaluation presented in the prior section. Indeed, even if users are interested in 10 km or coarser data rather than the 10 m resolution of the GEDI-S2 estimates, the ability to generate more recent global maps of tall and short crops is a considerable advantage of our approach compared to those relying on administrative data. We therefore provide both 10 m and 10 km aggregated versions of our data (see data availability statement).

\section{Discussion}\label{discussion}

\subsection{Sources of error} \label{errors}
\begin{figure}
	\centering
	\includegraphics[width=1.0\linewidth]{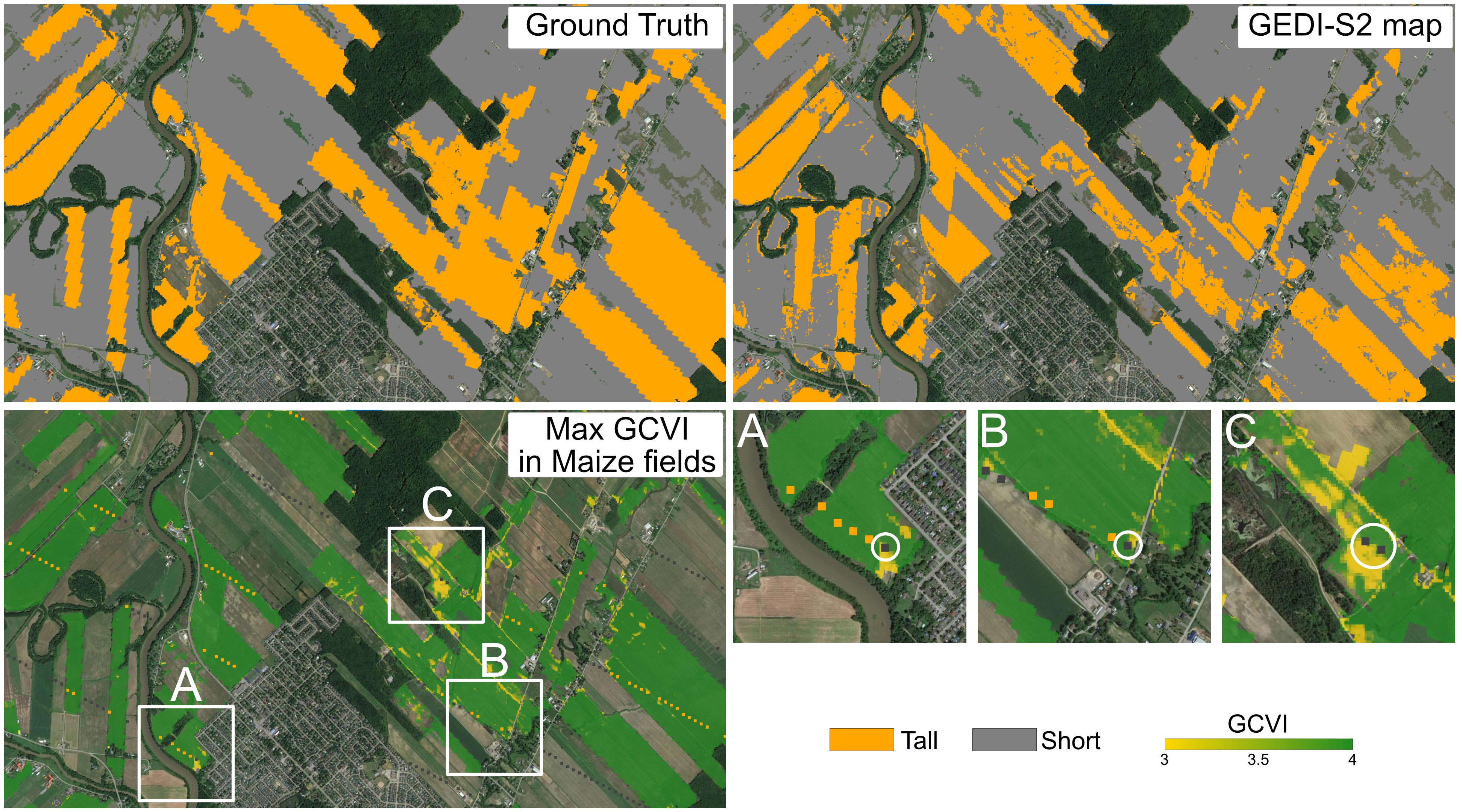}
    \caption{
    An example of GEDI-S2 errors in low biomass maize fields in Canada. Ground truth (top left) represents the Canada AAFC 2019 map. The corresponding GEDI-S2 2019 predictions (top right) miss several tall fields. S2 peak GCVI in the 2019 growing season (bottom left) indicates that omitted fields often have a peak GCVI below 4. A zoom into three examples (lower right) shows the individual GEDI shots and the predicted class (orange = tall, gray = short). Circles indicate examples where the shots were classified as short over low GCVI areas. Since the GEDI-S2 model is then trained on the predicted classes for these shots, it also incorrectly classifies some tall crop area as short.}
    \label{fig:lowVImap}
\end{figure}

Despite the overall encouraging performance of the GEDI-S2 approach, some regions show a clear underestimation of tall crop area. Visual inspection of the GEDI estimates used to train the GEDI-S2 models in these regions indicates that several of the training points are incorrectly labeled as short crops. An example for Canada illustrates this phenomenon well (Figure \ref{fig:lowVImap}). One likely cause of GEDI falsely predicting a short crop is that the biomass of the tall crop (in this case maize) is significantly less than the typical biomass in regions used to train the GEDI crop height model, resulting in a greater fraction of photon returns from close to the ground. One might argue that these maize fields should not, in fact, be considered a tall crop because they have insufficient biomass above the height of typical small crops. However, as the goal of this work is to reliably map crop types regardless of their biomass, it is important that all fields with maize be included in the same class. 

\begin{figure}
	\centering
	\includegraphics[width=0.7\linewidth]{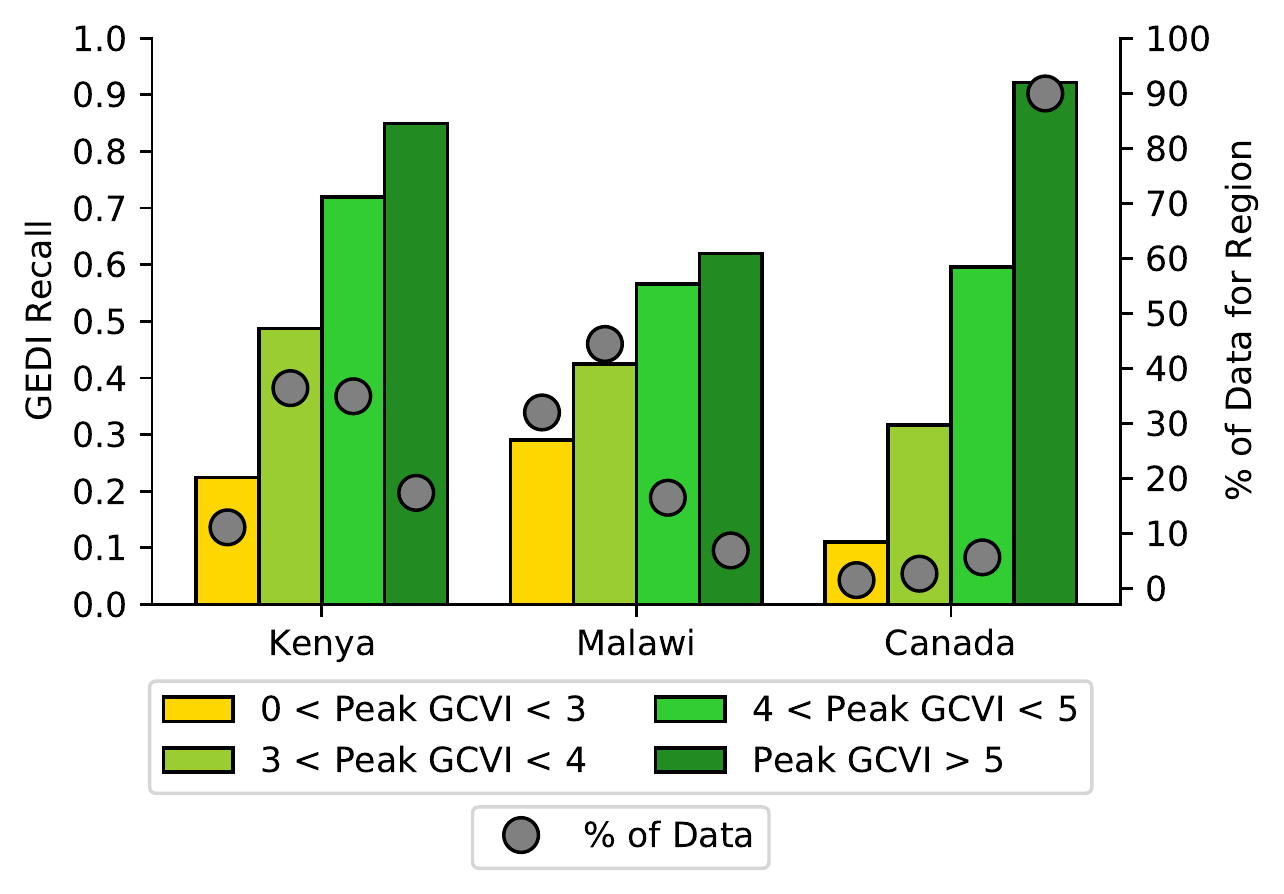}
    \caption{
The effect of peak GCVI on GEDI model recall for maize. Peak GCVI, a proxy for biomass, is computed from S2 time series as the maximum GCVI in the optimal month: July for Kenya, February for Malawi and August for Canada. Maize fields in the three regions are defined by respective reference maps: JRC 2021 Long Rains crop type map for Kenya,  2019 AAFC for Canada, and 2021 S2-Local map for Malawi. Recall is much higher on fields with peak GCVI above 4. In Kenya and Malawi, more than 75\% of fields have peak GCVI below 4 in these years.
    }
    \label{fig:lowVIbar}
\end{figure}

To further explore the hypothesis that GEDI struggles are related to low biomass of the tall crop, we consider three regions for which maize spans a range of low to high biomass -- Kenya, Malawi, and Canada. For each region, we take all GEDI shots that fall onto pixels predicted to be maize by either the local reference data (in the case of Kenya and Canada) or the S2-Local model (in the case of Malawi), and then split these GEDI shots into four groups based on the peak GCVI of the S2 pixel that overlaps the GEDI shot. Peak GCVI is used as a proxy for biomass, given that GCVI has been widely shown to correlate well with maize biomass \cite{gitelson2005remote}. We then calculate the fraction of maize GEDI shots predicted to be a tall crop. Consistent with our hypothesis, we find that the GEDI recall is much higher for fields with higher peak GCVI (Figure \ref{fig:lowVIbar}). Recall increases monotonically as the GCVI increases, and recall for pixels with a peak above 4 is at least double that for pixels with a peak VI below 3. Based on this analysis, we consider pixels with peak GCVI below 4 to be less reliable for GEDI-S2 predictions, and therefore provide a quality flag for these pixels in our final estimates. 
In general, the regions performing poorly in both the field-level and global evaluation are those for which peak VI is frequently below 4 (Figure \ref{fig:lowVImap},\ref{fig:lowVIbar}). A region like Canada still performs well overall because the frequency of fields with peak GCVI below 4 is very low (Figure \ref{fig:lowVIbar}). In contrast, more than half of maize fields in Kenya, Malawi, India, and many other regions have peak GCVI below this value, resulting in poor overall performance.

\subsection{Future improvements}
The strong agreement with independent reference data in many regions (Table \ref{table:validationTable}, Figure \ref{fig:validation}) indicates that the GEDI-S2 approach is a promising tool for global crop type mapping. Fully realizing its potential will require progress on several fronts, all of which are beyond the scope of the current paper but for which we anticipate progress is likely. First and foremost is the need to improve performance in areas with lower crop biomass, where the distinction between tall and short crops is reduced. One approach could be to retrain a separate GEDI model for these regions if enough high-quality reference data are available, although the small disparities between tall and short crop returns in these settings make that unlikely to work. Another approach could be to use semi-supervised methods where the GEDI shots predicted for high GCVI fields within low biomass regions are used as high quality labels for fine tuning \cite{weinstein2019individual,wu2017semi}. These semi-supervised approaches may also benefit from using S2 features that are less sensitive to peak biomass \cite{qiu2018mapping,lin2022early}. 

A second area for future work is to extend the GEDI-S2 approach to map multiple seasons in regions where more than one crop per year is typically grown. In the current work we focused only on the season with the highest proportion of tall crops, but many areas have two seasons that each have a significant fraction of tall crops (e.g., Eastern Africa, Northern India). Performing the GEDI-S2 training for multiple seasons would be a straightforward extension of the current work, with the main requirement likely being a shift in the window over which the S2 harmonic features are calculated.

A third extension could be to further discriminate among crops beyond the two categories of tall and short. With additional training data, it may be possible for GEDI returns to distinguish very short crops, such as legumes, from slightly taller crops such as rapeseed or cotton. For example, many areas in India are dominated in the monsoon season by rice, cotton, and sugarcane \cite{wang2020mapping}, each of which would fall into separate height classes. Even if the GEDI predictions from this approach are noisy, they could be effective labels for training a GEDI-S2 model. 

Alternatively, refining the crop classification could utilize complementary features that are unrelated to height. For example, distinguishing short winter or spring crops such as winter and barley from short summer crops is fairly simple based on phenological differences in optical or radar data \cite{veloso2017understanding,blickensdorfer2022mapping}, and rice can be accurately distinguished from other crops based on flooding patterns detected in radar imagery \cite{nelson2014towards,singha2019high}. Many agricultural landscapes will possess only one other main crop beyond wheat, maize, and rice, and so the ability to map these three could effectively map all major crop types (if the fourth crop is estimated as fields not in one of these three crops).  In much of North and South America, for example, maize and soybeans are by far the most common summer crops, and so a large fraction of non-maize area is in soybean. Further work is needed to test how far the tall vs. short crop distinction can help to solve the more general problem of mapping crop species.

\section{Conclusions}
\label{conclusions}

We sought in this study to test the general applicability of an approach that uses GEDI returns to train local crop type mapping models that use S2 data as input. Our general conclusion is that the approach exhibits considerable promise for advancing crop mapping. Tall and short crops were mapped with high accuracy in the majority of maize production systems, including most of the Americas, Europe, and East Asia. Specifically, we showed that GEDI returns can first be classified into tall and short crops, that the frequency of tall crops over time can be used to identify the appropriate months for S2 training (when tall crops are at their peak height), and that S2 models trained on these GEDI shots can accurately predict the GEDI crop height class in nearly all regions.  Only in rare cases, such as areas with high topographic variation, did S2 features fail to predict GEDI crop height class. We then showed that the predictions from the GEDI-S2 agree remarkably well with independent reference data at the field scale, as well as with broad patterns of tall crops in a much coarser global gridded dataset.

At the same time, we uncovered cases where the current implementation of GEDI-S2 is problematic. The most common cause for low accuracy appears to be low biomass of tall crops, which occurs frequently in Africa and South Asia. In these regions, the GEDI classification model consistently underestimated the frequency of tall crops. S2 models trained on these shots then inherit this under-prediction of tall crop area. Although this is a notable limitation of the current approach -- particularly because these regions are among those with the most limited ground data, and thus where an approach that relied on GEDI for training would be most valuable -- we anticipate that future work can greatly improve the performance in low biomass regions. Progress seems most likely for semi-supervised methods that can leverage the fact that even low biomass areas typically have a significant number of fields with high biomass that are accurately captured by GEDI.

\section*{Data availability statement}
Data will be made publicly available upon publication as assets in Google Earth Engine.

\section*{Acknowledgements}
We thank Rick Brandenburg for sharing the Malawi crop reference data, Marcel Schwieder for access to the Germany crop type maps, and the Google Earth Engine team for making large-scale computational resources available to researchers. This work was supported by the NASA Harvest Consortium (NASA Applied Sciences Grant No. 80NSSC17K0652, sub-award 54308-Z6059203 to DBL). SW was supported by the Ciriacy-Wantrup Postdoctoral Fellowship at the University of California, Berkeley. 


\section*{References}

\bibliographystyle{dcu}
\bibliography{bibl}

@article{luo2022developing,
  title={Developing High-Resolution Crop Maps for Major Crops in the European Union Based on Transductive Transfer Learning and Limited Ground Data},
  author={Luo, Yuchuan and Zhang, Zhao and Zhang, Liangliang and Han, Jichong and Cao, Juan and Zhang, Jing},
  journal={Remote Sensing},
  volume={14},
  number={8},
  pages={1809},
  year={2022},
  publisher={MDPI}
}

@article{kim2021review,
  title={A review of global gridded cropping system data products},
  author={Kim, Kwang-Hyung and Doi, Yasuhiro and Ramankutty, Navin and Iizumi, Toshichika},
  journal={Environmental Research Letters},
  volume={16},
  number={9},
  pages={093005},
  year={2021},
  publisher={IOP Publishing}
}

@article{you202110m,
	title = {The 10-m crop type maps in {Northeast} {China} during 2017–2019},
	volume = {8},
	copyright = {2021 The Author(s)},
	issn = {2052-4463},
	number = {1},
	journal = {Scientific Data},
	author = {You, Nanshan and Dong, Jinwei and Huang, Jianxi and Du, Guoming and Zhang, Geli and He, Yingli and Yang, Tong and Di, Yuanyuan and Xiao, Xiangming},
	month = feb,
	year = {2021},
	note = {Number: 1
Publisher: Nature Publishing Group},
	pages = {41}
}

@article{weinstein2019individual,
  title={Individual tree-crown detection in RGB imagery using semi-supervised deep learning neural networks},
  author={Weinstein, Ben G and Marconi, Sergio and Bohlman, Stephanie and Zare, Alina and White, Ethan},
  journal={Remote Sensing},
  volume={11},
  number={11},
  pages={1309},
  year={2019},
  publisher={MDPI}
}

@article{wu2017semi,
  title={Semi-supervised deep learning using pseudo labels for hyperspectral image classification},
  author={Wu, Hao and Prasad, Saurabh},
  journal={IEEE Transactions on Image Processing},
  volume={27},
  number={3},
  pages={1259--1270},
  year={2017},
  publisher={IEEE}
}

@article{dubayah2020global,
	title = {The {Global} {Ecosystem} {Dynamics} {Investigation}: {High}-resolution laser ranging of the {Earth}’s forests and topography},
	volume = {1},
	issn = {2666-0172},
	shorttitle = {The {Global} {Ecosystem} {Dynamics} {Investigation}},
	journal = {Science of Remote Sensing},
	author = {Dubayah, Ralph and Blair, James Bryan and Goetz, Scott and Fatoyinbo, Lola and Hansen, Matthew and Healey, Sean and Hofton, Michelle and Hurtt, George and Kellner, James and Luthcke, Scott and Armston, John and Tang, Hao and Duncanson, Laura and Hancock, Steven and Jantz, Patrick and Marselis, Suzanne and Patterson, Paul L. and Qi, Wenlu and Silva, Carlos},
	month = jun,
	year = {2020},
	pages = {100002}
}

@article{wang2019crop,
	title = {Crop type mapping without field-level labels: {Random} forest transfer and unsupervised clustering techniques},
	volume = {222},
	issn = {0034-4257},
	shorttitle = {Crop type mapping without field-level labels},
	journal = {Remote Sensing of Environment},
	author = {Wang, Sherrie and Azzari, George and Lobell, David B.},
	month = mar,
	year = {2019},
	pages = {303--317}
}

@article{qiu2018mapping,
  title={Mapping spatiotemporal dynamics of maize in China from 2005 to 2017 through designing leaf moisture based indicator from Normalized Multi-band Drought Index},
  author={Qiu, Bingwen and Huang, Yingze and Chen, Chongchen and Tang, Zhenghong and Zou, Fengli},
  journal={Computers and Electronics in Agriculture},
  volume={153},
  pages={82--93},
  year={2018},
  publisher={Elsevier}
}

@article{blickensdorfer2022mapping,
  title={Mapping of crop types and crop sequences with combined time series of Sentinel-1, Sentinel-2 and Landsat 8 data for Germany},
  author={Blickensd{\"o}rfer, Lukas and Schwieder, Marcel and Pflugmacher, Dirk and Nendel, Claas and Erasmi, Stefan and Hostert, Patrick},
  journal={Remote sensing of environment},
  volume={269},
  pages={112831},
  year={2022},
  publisher={Elsevier}
}

@article{han2021asiaricemap10m,
  title={AsiaRiceMap10m: High-resolution annual paddy rice maps for Southeast and Northeast Asia from 2017 to 2019},
  author={Han, Jichong and Zhang, Zhao and Luo, Yuchuan and Cao, Juan and Zhang, Liangliang and Cheng, Fei and Zhuang, Huimin and Zhang, Jing},
  journal={Earth Syst. Sci. Data Discuss},
  volume={211},
  pages={1--27},
  year={2021}
}

@article{kluger2021two,
	Author = {Dan M. Kluger and Sherrie Wang and David B. Lobell},
	Journal = {Remote Sensing of Environment},
	Pages = {112488},
	Title = {Two shifts for crop mapping: Leveraging aggregate crop statistics to improve satellite-based maps in new regions},
	Volume = {262},
	Year = {2021}}

@article{di2021combining,
  title={Combining GEDI and Sentinel-2 for wall-to-wall mapping of tall and short crops},
  author={Di Tommaso, Stefania and Wang, Sherrie and Lobell, David B},
  journal={Environmental Research Letters},
  volume={16},
  number={12},
  pages={125002},
  year={2021},
  publisher={IOP Publishing}
}

@article{potapov2021mapping,
  title={Mapping global forest canopy height through integration of GEDI and Landsat data},
  author={Potapov, Peter and Li, Xinyuan and Hernandez-Serna, Andres and Tyukavina, Alexandra and Hansen, Matthew C and Kommareddy, Anil and Pickens, Amy and Turubanova, Svetlana and Tang, Hao and Silva, Carlos Edibaldo and others},
  journal={Remote Sensing of Environment},
  volume={253},
  pages={112165},
  year={2021},
  publisher={Elsevier}
}

@article{potapov2022global,
  title={Global maps of cropland extent and change show accelerated cropland expansion in the twenty-first century},
  author={Potapov, Peter and Turubanova, Svetlana and Hansen, Matthew C and Tyukavina, Alexandra and Zalles, Viviana and Khan, Ahmad and Song, Xiao-Peng and Pickens, Amy and Shen, Quan and Cortez, Jocelyn},
  journal={Nature Food},
  volume={3},
  number={1},
  pages={19--28},
  year={2022},
  publisher={Nature Publishing Group}
}

@article{healey2020highly,
  title={Highly local model calibration with a new GEDI LiDAR asset on Google Earth Engine reduces landsat forest height signal saturation},
  author={Healey, Sean P and Yang, Zhiqiang and Gorelick, Noel and Ilyushchenko, Simon},
  journal={Remote Sensing},
  volume={12},
  number={17},
  pages={2840},
  year={2020},
  publisher={Multidisciplinary Digital Publishing Institute}
}

@article{begue2018remote,
  title={Remote sensing and cropping practices: A review},
  author={B{\'e}gu{\'e}, Agn{\`e}s and Arvor, Damien and Bellon, Beatriz and Betbeder, Julie and De Abelleyra, Diego and PD Ferraz, Rodrigo and Lebourgeois, Valentine and Lelong, Camille and Sim{\~o}es, Margareth and R. Ver{\'o}n, Santiago},
  journal={Remote Sensing},
  volume={10},
  number={1},
  pages={99},
  year={2018},
  publisher={MDPI}
}

@article{lin2022early,
  title={Early-and in-season crop type mapping without current-year ground truth: Generating labels from historical information via a topology-based approach},
  author={Lin, Chenxi and Zhong, Liheng and Song, Xiao-Peng and Dong, Jinwei and Lobell, David B and Jin, Zhenong},
  journal={Remote Sensing of Environment},
  volume={274},
  pages={112994},
  year={2022},
  publisher={Elsevier}
}

@article{nakalembe2021review,
  title={A review of satellite-based global agricultural monitoring systems available for Africa},
  author={Nakalembe, Catherine and Becker-Reshef, Inbal and Bonifacio, Rogerio and Hu, Guangxiao and Humber, Micheal Lawrence and Justice, Christina Jade and Keniston, John and Mwangi, Kenneth and Rembold, Felix and Shukla, Shraddhanand and others},
  journal={Global Food Security},
  volume={29},
  pages={100543},
  year={2021},
  publisher={Elsevier}
}

@article{veloso2017understanding,
  title={Understanding the temporal behavior of crops using Sentinel-1 and Sentinel-2-like data for agricultural applications},
  author={Veloso, Amanda and Mermoz, St{\'e}phane and Bouvet, Alexandre and Le Toan, Thuy and Planells, Milena and Dejoux, Jean-Fran{\c{c}}ois and Ceschia, Eric},
  journal={Remote Sensing of Environment},
  volume={199},
  pages={415--426},
  year={2017},
  publisher={Elsevier}
}

@article{boryan2011monitoring,
author = {Claire   Boryan  and  Zhengwei   Yang  and  Rick   Mueller  and  Mike   Craig},
title = {Monitoring US agriculture: the US Department of Agriculture, National Agricultural Statistics Service, Cropland Data Layer Program},
journal = {Geocarto International},
volume = {26},
number = {5},
pages = {341-358},
year  = {2011},
publisher = {Taylor & Francis},
doi = {10.1080/10106049.2011.562309}
}

@article{lobell2003remote,
  title={Remote sensing of regional crop production in the Yaqui Valley, Mexico: estimates and uncertainties},
  author={Lobell, David B and Asner, Gregory P and Ortiz-Monasterio, J Ivan and Benning, Tracy L},
  journal={Agriculture, Ecosystems \& Environment},
  volume={94},
  number={2},
  pages={205--220},
  year={2003},
  publisher={Elsevier}
}

@misc{france,
title={{Registre parcellaire graphique (RPG): contours des parcelles et îlots culturaux et leur groupe de cultures majoritaire}. {Available at https://www.data.gouv.fr/en/datasets/registre-parcellaire-graphique-rpg-contours-des-parcelles-et-ilots-culturaux-et-leur-groupe-de-cultures-majoritaire/} (accessed {2021-07-01}; verified {2021-07-01})},
author={{Agence de Services et de Paiement}},
year={2019},
}

@article{gitelson2005remote,
author = {Gitelson, Anatoly A. and Vina, Andres and Ciganda, Veronica and Rundquist, Donald C. and Arkebauer, Timothy J.},
title = {Remote estimation of canopy chlorophyll content in crops},
journal = {Geophysical Research Letters},
volume = {32},
number = {8},
pages = {},
year = {2005},
doi = {10.1029/2005GL022688}
}

@misc{aafc,
title={{Annual Crop Inventory [Online]}. {Available at https://open.canada.ca/data/en/dataset/ba2645d5-4458-414d-b196-6303ac06c1c9} (accessed {2021-07-01}; verified {2021-07-01})},
author={{Agriculture and Agri-Food Canada}},
year={2021},
institution={Agriculture and Agri-Food Canada}
}

@misc{aafcground,
    title={{Annual Crop Inventory Ground Truth Data [Online]}. {Available at https://https://open.canada.ca/data/en/dataset/503a3113-e435-49f4-850c-d70056788632} (accessed {2022-11-15}; verified {2022-11-15})},
    author={{Agriculture and Agri-Food Canada}},
    year={2021},
    institution={Agriculture and Agri-Food Canada}
}

@article{wang2020mapping,
	Author = {Wang, Sherrie and Di Tommaso, Stefania and Faulkner, Joey and Friedel, Thomas and Kennepohl, Alexander and Strey, Rob and Lobell, David B.},
	Journal = {Remote Sensing},
	Number = {18},
	Title = {Mapping Crop Types in Southeast India with Smartphone Crowdsourcing and Deep Learning},
	Volume = {12},
	Year = {2020}}

@article{singha2019high,
  title={High resolution paddy rice maps in cloud-prone Bangladesh and Northeast India using Sentinel-1 data},
  author={Singha, Mrinal and Dong, Jinwei and Zhang, Geli and Xiao, Xiangming},
  journal={Scientific data},
  volume={6},
  number={1},
  pages={1--10},
  year={2019},
  publisher={Nature Publishing Group}
}

@article{nelson2014towards,
  title={Towards an operational SAR-based rice monitoring system in Asia: Examples from 13 demonstration sites across Asia in the RIICE project},
  author={Nelson, Andrew and Setiyono, Tri and Rala, Arnel B and Quicho, Emma D and Raviz, Jeny V and Abonete, Prosperidad J and Maunahan, Aileen A and Garcia, Cornelia A and Bhatti, Hannah Zarah M and Villano, Lorena S and others},
  journal={Remote Sensing},
  volume={6},
  number={11},
  pages={10773--10812},
  year={2014},
  publisher={MDPI}
}

@article{zanaga2021esa,
  title={ESA WorldCover 10 m 2020 v100},
  author={Zanaga, Daniele and Van De Kerchove, Ruben and De Keersmaecker, Wanda and Souverijns, Niels and Brockmann, Carsten and Quast, Ralf and Wevers, Jan and Grosu, Alex and Paccini, Audrey and Vergnaud, Sylvain and others},
  journal={Zenodo: Geneve, Switzerland},
  year={2021}
}

@inproceedings{karra2021global,
  title={Global land use/land cover with Sentinel 2 and deep learning},
  author={Karra, Krishna and Kontgis, Caitlin and Statman-Weil, Zoe and Mazzariello, Joseph C and Mathis, Mark and Brumby, Steven P},
  booktitle={2021 IEEE international geoscience and remote sensing symposium IGARSS},
  pages={4704--4707},
  year={2021},
  organization={IEEE}
}

@article{fayad2020analysis,
  title={Analysis of GEDI elevation data accuracy for inland waterbodies altimetry},
  author={Fayad, Ibrahim and Baghdadi, Nicolas and Bailly, Jean St{\'e}phane and Frappart, Fr{\'e}d{\'e}ric and Zribi, Mehrez},
  journal={Remote Sensing},
  volume={12},
  number={17},
  pages={2714},
  year={2020},
  publisher={MDPI}
}

@article{fayad2022comparative,
  title={Comparative Analysis of GEDI’s Elevation Accuracy from the First and Second Data Product Releases over Inland Waterbodies},
  author={Fayad, Ibrahim and Baghdadi, Nicolas and Frappart, Fr{\'e}d{\'e}ric},
  journal={Remote Sensing},
  volume={14},
  number={2},
  pages={340},
  year={2022},
  publisher={MDPI}
}

@article{fayad2021terrain,
  title={Terrain slope effect on forest height and wood volume estimation from GEDI data},
  author={Fayad, Ibrahim and Baghdadi, Nicolas and Alcarde Alvares, Clayton and Stape, Jose Luiz and Bailly, Jean St{\'e}phane and Scolforo, Henrique Ferra{\c{c}}o and Cegatta, Italo Ramos and Zribi, Mehrez and Le Maire, Guerric},
  journal={Remote Sensing},
  volume={13},
  number={11},
  pages={2136},
  year={2021},
  publisher={MDPI}
}

@data{DVN/PRFF8V_2019,
    author = {{International Food Policy Research Institute}},
    publisher = {Harvard Dataverse},
    title = {{Global Spatially-Disaggregated Crop Production Statistics Data for 2010 Version 2.0}},
    year = {2019},
    version = {DRAFT VERSION},
    doi = {10.7910/DVN/PRFF8V},
    url = {https://doi.org/10.7910/DVN/PRFF8V}
}

@dataset{blickensdorfer_lukas_2021_5153047,
  author       = {Blickensdörfer, Lukas and
                  Schwieder, Marcel and
                  Pflugmacher, Dirk and
                  Nendel, Claas and
                  Erasmi, Stefan and
                  Hostert, Patrick},
  title        = {{National-scale crop type maps for Germany from 
                   combined time series of Sentinel-1, Sentinel-2 and
                   Landsat 8 data (2017, 2018 and 2019)}},
  month        = aug,
  year         = 2021,
  publisher    = {Zenodo},
  doi          = {10.5281/zenodo.5153047},
  url          = {https://doi.org/10.5281/zenodo.5153047}
}

@dataset{schwieder_marcel_2022_6451848,
  author       = {Schwieder, Marcel and
                  Erasmi, Stefan and
                  Nendel, Claas and
                  Hostert, Patrick},
  title        = {{National-scale crop type maps for Germany from 
                   combined time series of Sentinel-1, Sentinel-2 and
                   Landsat 8 data (2020)}},
  month        = apr,
  year         = 2022,
  publisher    = {Zenodo},
  doi          = {10.5281/zenodo.6451848},
  url          = {https://doi.org/10.5281/zenodo.6451848}
}

@article{lee2022mapping,
  title={Mapping Sugarcane in Central India with Smartphone Crowdsourcing},
  author={Lee, Ju Young and Wang, Sherrie and Figueroa, Anjuli Jain and Strey, Rob and Lobell, David B and Naylor, Rosamond L and Gorelick, Steven M},
  journal={Remote Sensing},
  volume={14},
  number={3},
  pages={703},
  year={2022},
  publisher={MDPI}
}

@article{schneider2021eurocrops,
  title={Eurocrops: A pan-european dataset for time series crop type classification},
  author={Schneider, Maja and Broszeit, Amelie and K{\"o}rner, Marco},
  journal={arXiv preprint arXiv:2106.08151},
  year={2021}
}

@misc{geoglam,
    title={Kenya AOI. European Commission, Joint Research Centre (JRC) [Dataset]. {Available at https://data.jrc.ec.europa.eu/dataset/5b6245d3-e561-4f6c-8c09-627888063d11} (accessed {2022-11-15}; verified {2022-11-15})},
    year={2021},
    author={{European Commission, Joint Research Centre (JRC)}}
}

@misc{ngws,
    title={Global land surface phenology - Number of growing seasons [Dataset]. {Available at http://data.europa.eu/89h/jrc-10112-10008} (accessed {2022-11-15}; verified {2022-11-15})},
    year={2018},
    author={{European Commission, Joint Research Centre (JRC)}}
}

@article{rembold2019asap,
  title={ASAP: A new global early warning system to detect anomaly hot spots of agricultural production for food security analysis},
  author={Rembold, Felix and Meroni, Michele and Urbano, Ferdinando and Csak, Gabor and Kerdiles, Herv{\'e} and Perez-Hoyos, Ana and Lemoine, Guido and Leo, Olivier and Negre, Thierry},
  journal={Agricultural systems},
  volume={168},
  pages={247--257},
  year={2019},
  publisher={Elsevier}
}

@article{gorelick2017google,
  title={Google Earth Engine: Planetary-scale geospatial analysis for everyone},
  author={Gorelick, Noel and Hancher, Matt and Dixon, Mike and Ilyushchenko, Simon and Thau, David and Moore, Rebecca},
  journal={Remote sensing of Environment},
  volume={202},
  pages={18--27},
  year={2017},
  publisher={Elsevier}
}

@misc{awesomegeedatasets,
    title={{samapriya/awesome-gee-community-datasets: Community Catalog (1.0.1)}. {Available at https://doi.org/10.5281/zenodo.7271726} (accessed {2022-11-15})},
    year={2022},
    author={Roy, Samapriya and Swetnam, Tyson and Robitaille, Alec and Trochim, Erin and Pasquarella, Valerie }
}

@article{grandini2020metrics,
  title={Metrics for multi-class classification: an overview},
  author={Grandini, Margherita and Bagli, Enrico and Visani, Giorgio},
  journal={arXiv preprint arXiv:2008.05756},
  year={2020}
}

@article{farr2007shuttle,
  title={The shuttle radar topography mission},
  author={Farr, Tom G and Rosen, Paul A and Caro, Edward and Crippen, Robert and Duren, Riley and Hensley, Scott and Kobrick, Michael and Paller, Mimi and Rodriguez, Ernesto and Roth, Ladislav and others},
  journal={Reviews of geophysics},
  volume={45},
  number={2},
  year={2007},
  publisher={Wiley Online Library}
}

@article{song2021evaluation,
  title={An evaluation of Landsat, Sentinel-2, Sentinel-1 and MODIS data for crop type mapping},
  author={Song, Xiao-Peng and Huang, Wenli and Hansen, Matthew C and Potapov, Peter},
  journal={Science of Remote Sensing},
  volume={3},
  pages={100018},
  year={2021},
  publisher={Elsevier}
}

@article{song2021massive,
  title={Massive soybean expansion in South America since 2000 and implications for conservation},
  author={Song, Xiao-Peng and Hansen, Matthew C and Potapov, Peter and Adusei, Bernard and Pickering, Jeffrey and Adami, Marcos and Lima, Andre and Zalles, Viviana and Stehman, Stephen V and Di Bella, Carlos M and others},
  journal={Nature Sustainability},
  volume={4},
  number={9},
  pages={784--792},
  year={2021},
  publisher={Nature Publishing Group}
}

\end{document}